\documentclass[12pt,english]{article}
\usepackage{subfigure}
\usepackage{graphicx}
\usepackage{rotating}

\usepackage{hyperref}
\usepackage{amssymb}

\providecommand{\tabularnewline}{\\}

\makeatletter

\usepackage{babel}
\makeatother
\usepackage{geometry}
\usepackage{babel}
\usepackage{graphics}
\def\beq{\begin{equation}}
\def\eeq{\end{equation}}
\def\Q{{\bf Q}}

\def\beqn{\begin{eqnarray}}
\def\eeqn{\end{eqnarray}}
\def\ba{\begin{eqnarray}}
\def\ea{\end{eqnarray}}
\def\ie{{\it i.e.}}
\def\eg{{\it e.g.}}
\def\half{{\textstyle{1\over 2}}}

\def\third{{\textstyle {1\over3}}}
\def\quarter{{\textstyle {1\over4}}}
\def\m{{\tt -}}

\def\p{{\tt +}}

\def\slash#1{#1\hskip-6pt/\hskip6pt}
\def\slk{\slash{k}}
\def\GeV{\,{\rm GeV}}
\def\TeV{\,{\rm TeV}}
\def\y{\,{\rm y}}

\newcommand{\beqa}{\begin{eqnarray}}
\newcommand{\eeqa}{\end{eqnarray}}
\newcommand{\eps}{\epsilon}

\makeatletter

\providecommand{\LyX}{L\kern-.1667em\lower.25em\hbox{Y}\kern-.125emX\@}

\makeatother

\def\inbar{\,\vrule height1.5ex width.4pt depth0pt}

\def\IC{\relax\hbox{$\inbar\kern-.3em{\rm C}$}}
\def\IQ{\relax\hbox{$\inbar\kern-.3em{\rm Q}$}}
\def\IR{\relax{\rm I\kern-.18em R}}
 \font\cmss=cmss10 \font\cmsss=cmss10 at 7pt

\def\IZ{\relax\ifmmode\mathchoice
 {\hbox{\cmss Z\kern-.4em Z}}{\hbox{\cmss Z\kern-.4em Z}}
 {\lower.9pt\hbox{\cmsss Z\kern-.4em Z}}
 {\lower1.2pt\hbox{\cmsss Z\kern-.4em Z}}\else{\cmss Z\kern-.4em Z}\fi}

\begin{document}

\begin{center}
\vspace{1.5cm}
{\Large \bf  Searching for Extra $Z^\prime$ from
Strings and Other Models at the LHC with Leptoproduction\\}

\vspace{1cm}
{\bf Claudio Corian\`{o} $^{a,b }$,
	Alon E. Faraggi ${^c}$ and
	Marco Guzzi${^a}$ }

\vspace{1cm}

{\it  $^a$Dipartimento di Fisica, Universit\`{a} del Salento, and \\
 INFN Sezione di Lecce,  Via Arnesano 73100 Lecce, Italy}\\
\vspace{.5cm}
{\it $^b$ Department of Physics and Institute of Plasma Physics \\
University of Crete, 71003 Heraklion, Greece}\\
\vspace{.5cm}
{\it
$^c$Department of Mathematical Sciences,\\
 University of Liverpool, Liverpool L69 7ZL, United Kingdom\\}
\vspace{.12in}
\begin{abstract}
Discovery potentials for extra neutral interactions at the Large Hadron
Collider in forthcoming experiments are analyzed using resonant
leptoproduction. For this purpose we use high precision
next-to-next-to-leading order (NNLO) determinations
of the QCD background in this channel, at the tail of the Drell-Yan
distributions, in the invariant mass region around $ 0.8< Q < 2.5 $ TeV.
We focus our analysis primarily on a novel string-inspired $Z^\prime$,
obtained in left-right symmetric free fermionic heterotic string models
and whose existence at low energies is motivated by its role in
suppressing proton decay mediation. 
We analyze the parametric dependence of the predictions and perform
comparison with other models based on bottom up approaches,
that are constructed by requiring anomaly cancellation and enlarged Higgs structure.
We show that the results are not
particularly sensitive to the specific charge assignments. This may
render quite difficult the extraction of significant information from the
forward-backward asymmetries on the resonance, assuming that these are
possible due to a sizeable width. The challenge to discover extra (non
anomalous) $Z^\prime$ in this kinematic region remains strongly
dependent on the size of the new gauge coupling. Weakly coupled extra
$Z^\prime$ will not be easy to identify  even with a very good
theoretical determination of the QCD background through NNLO.
\end{abstract}
\end{center}
\newpage

\setcounter{page}{1}

\def\at{ }
\def\beq{\begin{equation}}
\def\eeq{\end{equation}}
\def\beqn{\begin{eqnarray}}
\def\eeqn{\end{eqnarray}}
\def\no{\noindent }
\def\nolabel{\nonumber }
\def\ba{\begin{eqnarray}}
\def\ea{\end{eqnarray}}

\def\NA{non--Abelian }

\def\gsim{{\buildrel >\over \sim}}
\def\lsim{{\buildrel <\over \sim}}

\def\ie{i.e., }
\def\eg{{\it e.g.}\ }
\def\eq#1{eq.\ (\ref{#1})}

\def\lt{<}

\def\slash#1{#1\hskip-6pt/\hskip6pt}
\def\slk{\slash{k}}

\def\dag{\dagger}
\def\qandq{\quad {\rm and} \quad}
\def\qand{\quad {\rm and} }
\def\andq{ {\rm and} \quad }
\def\qwithq{\quad {\rm with} \quad}
\def\qwith{ \quad {\rm with} }
\def\withq{ {\rm with} \quad}

\def\fhalf{\frac{1}{2}}
\def\fsqrt{\frac{1}{\sqrt{2}}}
\def\half{{\textstyle{1\over 2}}}
\def\third{{\textstyle {1\over3}}}
\def\quarter{{\textstyle {1\over4}}}
\def\sixth{{\textstyle {1\over6}}}
\def\m{$\phantom{-}$}
\def\j{$-$}
\def\ps{{\tt +}}
\def\pps{\phantom{+}}

\def\Tr{{\rm Tr}\, }
\def\tr{{\rm tr}\, }

\def\MP{M_{P}}
\def\GeV{\,{\rm GeV}}
\def\TeV{\,{\rm TeV}}

\def\lam#1{\lambda_{#1}}
\def\non{\nonumber}
\def\smgg{ $SU(3)_C\times SU(2)_L\times U(1)_Y$ }
\def\smggb{ $SU(3)_C\times SU(2)_L\times U(1)_Y$}
\def\SM{Standard--Model }
\def\SUSY{supersymmetry }
\def\SSSM{supersymmetric standard model}
\def\MSSM{minimal supersymmetric standard model}
\def\MSSSM{MS$_{str}$SM }
\def\MSSSMc{MS$_{str}$SM, }
\def\obs{{\rm observable}}
\def\sig{{\rm singlets}}
\def\hid{{\rm hidden}}
\def\MS{M_{str}}
\def\Ms{$M_{str}$}
\def\MP{M_{P}}

\def\vev#1{\langle #1\rangle}
\def\mvev#1{|\langle #1\rangle|^2}

\def\UA{U(1)_{\rm A}}
\def\QA{Q^{(\rm A)}}
\def\mssm{SU(3)_C\times SU(2)_L\times U(1)_Y} 

\def\KM{Ka\v c--Moody }

\def\y{\,{\rm y}}
\def\l{\langle}
\def\r{\rangle}
\def\o#1{\frac{1}{#1}}

\def\zi{z_{\infty}}

\def\hb#1{\bar{h}_{#1}}
\def\Htw{{\tilde H}}
\def\chibar{{\overline{\chi}}}
\def\qbar{{\overline{q}}}
\def\ibar{{\overline{\imath}}}
\def\jbar{{\overline{\jmath}}}
\def\Hbar{{\overline{H}}}
\def\Qbar{{\overline{Q}}}
\def\abar{{\overline{a}}}
\def\alphabar{{\overline{\alpha}}}
\def\betabar{{\overline{\beta}}}
\def\tautwo{{ \tau_2 }}
\def\thetatwo{{ \vartheta_2 }}
\def\thetathree{{ \vartheta_3 }}
\def\thetafour{{ \vartheta_4 }}
\def\ttwo{{\vartheta_2}}
\def\tthree{{\vartheta_3}}
\def\tfour{{\vartheta_4}}
\def\ti{{\vartheta_i}}
\def\tj{{\vartheta_j}}
\def\tk{{\vartheta_k}}
\def\calF{{\cal F}}
\def\smallmatrix#1#2#3#4{{ {{#1}~{#2}\choose{#3}~{#4}} }}
\def\ab{{\alpha\beta}}
\def\Minv{{ (M^{-1}_\ab)_{ij} }}
\def\ii{{(i)}}
\def\V{{\bf V}}
\def\N{{\bf N}}

\def\b{{\bf b}}
\def\S{{\bf S}}
\def\X{{\bf X}}
\def\I{{\bf I}}
\def\bone{{\mathbf 1}}
\def\bo{{\mathbf 0}}
\def\bs{{\mathbf S}}
\def\mS{{\mathbf S}}
\def\bS{{\mathbf S}}
\def\bb{{\mathbf b}}
\def\mb{{\mathbf b}}
\def\mX{{\mathbf X}}
\def\mY{{\mathbf Y}}
\def\bX{{\mathbf X}}
\def\mI{{\mathbf I}}
\def\bI{{\mathbf I}}
\def\balpha{{\mathbf \alpha}}
\def\bbeta{{\mathbf \beta}}
\def\bgamma{{\mathbf \gamma}}
\def\bxi{{\mathbf \xi}}
\def\malpha{{\mathbf \alpha}}
\def\mbeta{{\mathbf \beta}}
\def\mgamma{{\mathbf \gamma}}
\def\mzeta{{\mathbf \zeta}}
\def\mxi{{\mathbf \xi}}
\def\bphi{\overline{\Phi}}

\def\eps{\epsilon}

\def\t#1#2{{ \Theta\left\lbrack \matrix{ {#1}\cr {#2}\cr }\right\rbrack }}
\def\C#1#2{{ C\left\lbrack \matrix{ {#1}\cr {#2}\cr }\right\rbrack }}
\def\tp#1#2{{ \Theta'\left\lbrack \matrix{ {#1}\cr {#2}\cr }\right\rbrack }}
\def\tpp#1#2{{ \Theta''\left\lbrack \matrix{ {#1}\cr {#2}\cr }\right\rbrack }}
\def\l{\langle}
\def\r{\rangle}

\def\op#1{$\Phi_{#1}$}
\def\opp#1{$\Phi^{'}_{#1}$}
\def\opb#1{$\overline{\Phi}_{#1}$}
\def\opbp#1{$\overline{\Phi}^{'}_{#1}$}
\def\oppb#1{$\overline{\Phi}^{'}_{#1}$}
\def\oppx#1{$\Phi^{(')}_{#1}$}
\def\opbpx#1{$\overline{\Phi}^{(')}_{#1}$}

\def\oh#1{$h_{#1}$}
\def\ohb#1{${\bar{h}}_{#1}$}
\def\ohp#1{$h^{'}_{#1}$}

\def\oQ#1{$Q_{#1}$}
\def\odc#1{$d^{c}_{#1}$}
\def\ouc#1{$u^{c}_{#1}$}

\def\oL#1{$L_{#1}$}
\def\oec#1{$e^{c}_{#1}$}
\def\oNc#1{$N^{c}_{#1}$}

\def\oH#1{$H_{#1}$}
\def\oV#1{$V_{#1}$}
\def\oHs#1{$H^{s}_{#1}$}
\def\oVs#1{$V^{s}_{#1}$}

\def\p#1{{\Phi_{#1}}}
\def\pp#1{{\Phi^{'}_{#1}}}
\def\pb#1{{{\overline{\Phi}}_{#1}}}
\def\pbp#1{{{\overline{\Phi}}^{'}_{#1}}}
\def\ppb#1{{{\overline{\Phi}}^{'}_{#1}}}
\def\ppx#1{{\Phi^{(')}_{#1}}}
\def\pbpx#1{{\overline{\Phi}^{(')}_{#1}}}

\def\h#1{h_{#1}}
\def\hb#1{{\bar{h}}_{#1}}
\def\hp#1{h^{'}_{#1}}

\def\Q#1{Q_{#1}}
\def\dc#1{d^{c}_{#1}}
\def\uc#1{u^{c}_{#1}}

\def\L#1{L_{#1}}
\def\ec#1{e^{c}_{#1}}
\def\Nc#1{N^{c}_{#1}}

\def\H#1{H_{#1}}
\def\V#1{V_{#1}}
\def\Hs#1{{H^{s}_{#1}}}
\def\Vs#1{{V^{s}_{#1}}}

\def\fdtv{FD2V }
\def\fdtp{FD2$^{'}$ }
\def\fdtpv{FD2$^{'}$v }

\def\FD2pv{FD2$^{'}$V }
\def\FD2p{FD2$^{'}$ }

\def\AEF{A.E. Faraggi}
\def\AP#1#2#3{{\it Ann.\ Phys.}\/ {\bf#1} (#2) #3}
\def\EPJ#1#2#3{{\it The Eur.\ Phys.\ Jour.\/} {\bf C#1} (#2) #3}
\def\NPB#1#2#3{{\it Nucl.\ Phys.}\/ {\bf B#1} (#2) #3}
\def\NPBPS#1#2#3{{\it Nucl.\ Phys.}\/ {{\bf B} (Proc. Suppl.) {\bf #1}} (#2) 
 #3}
\def\PLB#1#2#3{{\it Phys.\ Lett.}\/ {\bf B#1} (#2) #3}
\def\PRD#1#2#3{{\it Phys.\ Rev.}\/ {\bf D#1} (#2) #3}
\def\PRL#1#2#3{{\it Phys.\ Rev.\ Lett.}\/ {\bf #1} (#2) #3}
\def\PRT#1#2#3{{\it Phys.\ Rep.}\/ {\bf#1} (#2) #3}
\def\PTP#1#2#3{{\it Prog.\ Theo.\ Phys.}\/ {\bf#1} (#2) #3}
\def\MODA#1#2#3{{\it Mod.\ Phys.\ Lett.}\/ {\bf A#1} (#2) #3}
\def\IJMP#1#2#3{{\it Int.\ J.\ Mod.\ Phys.}\/ {\bf A#1} (#2) #3}
\def\nuvc#1#2#3{{\it Nuovo Cimento}\/ {\bf #1A} (#2) #3}
\def\JHEP#1#2#3{{JHEP} {\textbf #1}, (#2) #3}
\def\RPP#1#2#3{{\it Rept.\ Prog.\ Phys.}\/ {\bf #1} (#2) #3}
\def\etal{{\it et al.\/}}
\def\SJNP#1#2#3{{\it Sov.\ J.\ Nucl.\ Phys.}\/ {\bf #1} (#2) #3}
\def\YF#1#2#3{{\it Yad.\ Fiz.}\/ {\bf #1} (#2) #3}

\hyphenation{su-per-sym-met-ric non-su-per-sym-met-ric}
\hyphenation{space-time-super-sym-met-ric}
\hyphenation{mod-u-lar mod-u-lar--in-var-i-ant}


\section{Introduction} 

The search for neutral currents mediated by extra gauge bosons $(Z^\prime)$
at the Large Hadron Collider will gather considerable attention in the
next few years \cite{reviews}. Additional Abelian gauge interactions
arise frequently in many extensions of the Standard Model, like in
left--right symmetric models, in Grand Unified
Theories (GUTs) and in string inspired constructions \cite{reviews}.
It has also been suggested that the existence of a low scale $Z^\prime$
may account for the suppression of proton decay mediating operators in
supersymmetric theories and otherwise \cite{pszp, pati, cfg}.
Abelian gauge structures may also play a 
considerable role in fixing the structure of the flavor sector, for
instance in pinning down the neutrinos mass matrix.
Anomaly cancellation conditions, when supported also by an extended Higgs
and fermion family structure - for instance by the inclusion of 
right-handed neutrinos - may allow non-sequential solutions (i.e. charge 
assignments which are not proportional to the hypercharge) 
that are phenomenologically interesting and could be studied
by ATLAS and CMS. Furthermore, within left--right 
symmetric models, and their underlying $SO(10)$ embedding, the global 
baryon minus lepton number $(B-L)$ of the Standard Model is
promoted to a local symmetry.
Abelian gauge extensions are therefore among the
most well motivated extensions of the Standard Model.
For these reasons, the identification of the origin of the extra neutral
interaction in future collider experiments will be an important and
challenging task. 
In particular, measurements of the charge asymmetries - both for
the rapidity distributions and for the related total cross section - and 
of the forward-backward asymmetries, may be a way to gather information 
about the structure of these new neutral currents interactions, although 
in the models that we have studied this looks pretty difficult, given the 
low statistics.

As an extra Z$^\prime$ is common in model
building, the differences among the various constructions may remain
unresolved, unless additional physical requirements are imposed on these
models in order to strengthen the possibility for their unique
identification. In this work we analyze the potential for the discovery
of an extra Z$^\prime$ arising in a specific string
construction, which is motivated not only by an anomaly-free structure, as 
in most of the bottom--up models considered in the previous literature,
but with some additional requirements coming from an adequate suppression
of proton decay mediation. Bottom up approaches based only on anomaly 
cancellation are, in this respect, less constraining compared to models
derived either from a string construction or
from theories of grand unification (GUTs) and can only provide
a basic framework within which to direct the experimental
searches. At the same time the search for extra neutral interactions has
to proceed in some generality and be unbiased, looking for resonances in
several complementary channels.  In this work we will investigate
the relation between more constrained and less constrained searches of
extra neutral gauge bosons by choosing as a channel leptoproduction and
proceed with a comparison of some proposals that have been presented in the
recent literature. Our main interest is focused around an extra
Z$^\prime$ which has been derived using the free fermionic formulation 
of string theory in a specific class of left--right symmetric string models.
The new abelian structure is determined not 
just as an attempt to satisfy some additional physical requirements, on 
which we elaborate below, but is naturally derived from a class of string 
models which have been extensively studied in detail in the past 
two decades \cite{fsu5,alr,slm,cfs}. 

Our paper is organized as follows: in section \ref{hszp} we
discuss the origin of $Z^\prime$ in heterotic--string models.
We discuss in some details the origin of the charge assignment
under the $Z^\prime$, which is motivated from proton decay considerations
and differs from those that have traditionally been discussed in the 
literature. Then we move to define the conventions in regard to the charge
assignments and the Higgs structure of the models that we consider, which
are characterized by a gauge structure which enlarges the gauge group of
the Standard Model by one extra $U(1)$. Our numerical analysis of the
invariant mass distributions for leptoproduction is performed by
varying both the coupling of the extra $U(1)$ and the mass of the new
gauge boson. The dependence on these parameters of the models that we
discuss are studied rather carefully in a kinematic region which can be
accessed at the LHC.  We compare these results with those obtained for a
group of 4 different models, introduced in \cite{Carena}, for which we
perform a similar analysis using leptoproduction. From this analysis it
is quite evident that the search for extra neutral currents at the LHC
is a rather difficult enterprise in  leptoproduction, unless the coupling
of the new gauge interaction is quite sizeable.
 
\section{Heterotic--string inspired $Z^\prime$}\label{hszp}

Phenomenological string models can be built in the heterotic--string or, using
brane constructions, in the type I string. The advantage of the former is that
it produces states in spinorial representations of the gauge group,
and hence allows for the $SO(10)$ embedding of the matter spectrum. 
The ten dimensional supersymmetric heterotic--string vacua 
give rise to effective field theories
that descend from the $E_8\times E_8$ or $SO(32)$ gauge groups. 
The first case gives rise to additional $Z^\prime$s
that arise in the $SO(10)$ and $E_6$ extensions of the Standard Model, 
and are the cases mostly studied in the literature \cite{reviews}.
A basis for the extra $Z^\prime$ arising in these models is
formed by the two groups $U(1)_\chi$ and $U(1)_\psi$ via the decomposition
$E_6\rightarrow SO(10)\times U(1)_\psi$ and $SO(10)\rightarrow SU(5)\times
U(1)_\chi$ \cite{reviews}.
Additional, flavor non--universal $U(1)$'s, may arise in
heterotic $E_8\times E_8$ string models from the $U(1)$ currents in the Cartan
subalgebra of the four dimensional gauge group, that are external to $E_6$.
Non--universal $Z^\prime$s typically must be beyond the LHC reach, to
avoid conflict with Flavor Changing Neutral Currents (FCNC) constraints.
Recently \cite{cfg} a novel $Z^\prime$ in quasi--realistic
string models that do not descend from the heterotic 
$E_8\times E_8$ string has been identified. Under the new $U(1)$
symmetry left--handed components and right--handed components in
the 16 spinorial $SO(10)$ representation, of each Standard Model generation,
have charge $-1/2$ and $+1/2$, respectively. As a result, the extra $U(1)$ is
family universal and anomaly free. It arises in left-right symmetric
string models \cite{cfs}, in which the $SO(10)$ symmetry is broken directly
at the string level to $SU(3)\times U(1)_{B-L}\times SU(2)_L\times
SU(2)_R\times U(1)_{Z^\prime}\times U(1)^n\times {\rm hidden}$
\cite{cfs}. The $U(1)^n$ are flavor dependent $U(1)$s that are
broken near the string scale. The Standard Model matter states are
neutral under the hidden sector gauge group, which in these
string models is typically a rank eight group. It is important to
note that the fact that the spectrum is derived from a string vacuum
that satisfies the modular invariance constraints, establishes
that the model is free from gauge and gravitational anomalies.
The pattern of $U(1)_{Z^\prime}$
charges in the quasi--realistic string models of ref. \cite{cfs}
does not arise in related string models in which the $SO(10)$ symmetry
is broken to the $SU(5)\times U(1)$ \cite{fsu5}, the $SO(6)\times SO(4)$
\cite{alr}, or $SU(3)\times SU(2)\times U(1)^2$ \cite{slm},
subgroups. The reason for
the distinction of the left--right symmetric string models is the
boundary condition assignment to the world--sheet free fermions 
that generate the $SO(10)$ symmetry in the basis vectors that break the 
$SO(10)$ symmetry to one of its subgroups. The world--sheet fermions that 
generate the rank eight observable gauge group in the free fermionic models
are denoted by $\{{\bar\psi}^{1,\cdots,5},{\bar\eta}^{1,2,3}\}$, where 
${\bar\psi}^{1,\cdots,5}$ generate an $SO(10)$ symmetry, and 
${\bar\eta}^{1,2,3}$ produce three $U(1)$ currents\footnote{for
reviews and the notation used in free fermionic string models see {\it e.g.}
\cite{ffmreviews} and references therein.}. Additional
observable gauged $U(1)$ currents may arise at enhanced symmetry 
points of the compactified six dimensional lattice. 
The $SO(10)$ gauge group is broken to one of its subgroups
$SU(5)\times U(1)$, $SO(6)\times SO(4)$ or
$SU(3)\times SU(2)\times U(1)^2$ by the assignment of
boundary conditions to the set ${\bar\psi}^{1\cdots5}_{1\over2}$:

\beqn
&1.&b\{{{\bar\psi}^{1\cdots5}}{\bar\eta}^{1,2,3}\}=
\{
{1\over2}{1\over2}{1\over2}{1\over2}{1\over2}{1\over2}{1\over2}{1\over2}
\}
\Rightarrow SU(5)\times U(1)\times U(1)^3,\label{su51so64breakingbc}\\
&2.&b\{{{\bar\psi}^{1\cdots5}}{\bar\eta}^{1,2,3}\}=
\{1 1 1 0 0 0 0 0 \}~~~~~
  \Rightarrow SO(6)\times SO(4)\times U(1)^3.\nonumber
\eeqn

To break the $SO(10)$ symmetry
to\footnote{$U(1)_C= {3\over2}U(1)_{B-L}; U(1)_L=2U(1)_{T_{3_R}}.$}
$SU(3)_C\times SU(2)_L\times U(1)_C\times U(1)_L$
both steps, 1 and 2, are used, in two separate basis vectors.
The breaking pattern
$SO(10)\rightarrow SU(3)_C\times SU(2)_L\times SU(2)_R \times U(1)_{B-L}$
is achieved by the following assignment in two separate basis
vectors
\beqn
&1.&b\{{{\bar\psi}^{1\cdots5}}{\bar\eta}^{1,2,3}\}=
\{1 1 1 0 0 0 0 0\}~~~
  \Rightarrow SO(6)\times SO(4)\times U(1)^3,\label{su3122breakingbc}\\
&2.&b\{{{\bar\psi}^{1\cdots5}}{\bar\eta}^{1,2,3}\}=
\{{1\over2}{1\over2}{1\over2}00{1\over2}{1\over2}{1\over2}
\}\Rightarrow SU(3)_C\times U(1)_C
\times SU(2)_L\times SU(2)_R\times U(1)^3\nonumber
\eeqn

The distinction between the symmetry breaking patterns in eq.
(\ref{su51so64breakingbc}) and eq. (\ref{su3122breakingbc})
is with respect to the charges of the Standard Model states under the three
flavor dependent $U(1)$ symmetries $U(1)_{1,2,3}$ that arise from the three
world--sheet fermions ${\bar\eta}^{1,2,3}$. In the free fermionic models, the
states of each Standard Model generation fit
into the 16 representation of $SO(10)$, and are charged with respect to one
of the three flavor $U(1)$ symmetries. For the symmetry breaking pattern
given in eq. (\ref{su51so64breakingbc}) the charge is always $+1/2$, {\it i.e.}
\beq
Q_j
\left(16 = \{Q, L, U, D, E, N\}\right) ~=~ + {1\over2}
\label{su51so64u1jcharges}
\eeq
whereas for the symmetry breaking pattern in eq. (\ref{su3122breakingbc})
the charges are

\beqn
Q_j (Q_L, L_L) & = & - {1\over2} \nonumber\\
Q_j (Q_R = \{U,D\}, L_R=\{E,N\}) & = & + {1\over2} 
\label{su3122charges}
\eeqn
As a result in the models admitting the symmetry breaking pattern 
eq. (\ref{su51so64breakingbc}) the combination 
\beq
U(1)_\zeta = U(1)_1 + U(1)_2 + U(1)_3.
\label{u1zeta}
\eeq
is anomalous, whereas in the models admitting the symmetry breaking pattern 
(\ref{su3122breakingbc}) it is anomaly free. The distinction between
the two boundary condition assignments given in eqs. 
(\ref{su51so64breakingbc}) and (\ref{su3122breakingbc}), 
and the consequent symmetry breaking patterns, is important
for the following reason. 
Whereas the first is obtained from an $N=4$ vacuum with 
$E_8\times E_8$ or $SO(16)\times SO(16)$ gauge symmetry, arising from the
$\{{\bar\psi}^{1,\cdots,5},{\bar\eta}^{1,2,3}{\bar\phi}^{1,\cdots,8}\}$
world--sheet fermions, which generate the observable and hidden sectors
gauge symmetries, the second cannot be obtained from these $N=4$ vacua,
but rather from an $N=4$ vacuum with $SO(16)\times E_7\times E_7$ gauge
symmetry, where we have included here also the symmetry arising from the
compactified lattice at the enhanced symmetry point. The important fact from
the point of view of the $Z^\prime$ phenomenology in which we are interested
is that the first case gives rise to the type of string inspired $Z^\prime$
that arises in models with an underlying $E_6$ symmetry. Whereas the $E_6$
may be broken at the string level, rather than in the effective low energy
field theory, the crucial point is that the charge assignment of the Standard
Model states is fixed by the underlying $E_6$ symmetry. The entire literature
on string inspired $Z^\prime$ studies this type of $E_6$ inspired $Z^\prime$.
The second class, however, is novel and has not been studied in the literature.
In this respect it would be interesting to examine how the symmetry breaking
pattern (\ref{su3122breakingbc}) and the corresponding charge assignments
(\ref{su3122charges}) can be obtained in heterotic orbifold models in which
one starts from a ten dimensional theory and compactifies to four dimensions,
rather than starting directly with a theory in four dimensions, as is done in
the free fermionic models. This understanding may highlight the relevance of
ten dimensional backgrounds that have thus far been ignored in the literature.
{}From the point of view of the $Z^\prime$ phenomenology, which is our
interest here, the crucial point will be to resolve between the different
$Z^\prime$ models and the fermion charges, which will reveal the relevance
of a particular symmetry breaking pattern.

The existence of the
extra $Z^\prime$ at low energies, within reach of the LHC,
is motivated by proton longevity,
and the suppression of the proton decay mediating operators
\cite{pszp, pati, cfg}.
The important property of this $Z^\prime$ is that it forbids dimension
four, five and six proton decay mediating operators.
The extra $U(1)$ is anomaly free and family universal. It allows
the fermions Yukawa couplings to the Higgs field and the
generation of small neutrino masses via a seesaw mechanism.
String models contain several $U(1)$ symmetries that suppress
the proton decay mediating operators \cite{pati}. However, these
are typically non--family universal. They constrain the fermion
mass terms and hence must be broken at a high scale.
Thus, the existence of a $U(1)$ symmetry that can remain unbroken
down to low energies is highly nontrivial. The $U(1)$ symmetry
in ref. \cite{cfs, cfg} satisfies all of these requirements.
Furthermore, as the generation of small neutrino masses in the
string models arises from the breaking of the $B-L$ current,
the extra $U(1)$ allows lepton number violating terms, but forbids
the baryon number violating terms. Hence, it predicts that
$R$--parity is violated and its phenomenological implications
for SUSY collider searches differ substantially from models
in which $R$--parity is preserved. The charges of the
Standard Model states under the $Z^\prime$ are displayed in table \ref{table1}.
Also displayed in the table are the charges under 
$U(1)_{\zeta^\prime}=U_C-U_L$, which is the Abelian combination of the 
Cartan generators of the underlying $SO(10)$ symmery that is orthogonal
to the weak hypercharge $U(1)_Y$. The charges under the $U(1)$ combination 
given in eq. (\ref{u1zeta}) are displayed in table \ref{table1} as well. 
These two $U(1)$'s are broken by the VEV that induces the seesaw
mechanism, and the combination 
\beq
U(1)_{Z^\prime}= {1\over 5} U(1)_{\zeta^\prime}-U(1)_\zeta
\eeq
is left unbroken down to low energies in order to suppress the proton decay
mediating operators. The charges of the Standard Model states under this 
$U(1)_{Z^\prime}$ are displayed in table \ref{table1}. 

\section{The interactions for $U(1)_{Z^\prime}$}
In this section we fix our conventions and describe the structure of the 
new neutral sector that we are going to analyze numerically in
leptoproduction afterwards. The notations are the same both in the case 
of the string model and for the other models that we will investigate. We
show in (\ref{table1}) the field content of the string model obtained 
within the free fermionic construction discussed above. Of the 3 extra
$U(1)$, we will decouple the two gauge bosons  denoted by $\zeta$,
$\zeta'$ and keep only the $Z^\prime$.  The assumption of decoupling of these
extra components are realistic if they are massive enough  ($> 5$ TeV) so
to neglect their influence on the lowest new resonance. We have chosen a
mass $M_{Z^\prime}$ around $0.8$ TeV. We recall that a reasonable region 
where the new extra gauge boson have a chance of being detected is below
the 5 TeV range.

The fermion-fermion-$Z^{\prime}$ interaction  is given by
\ba
\sum_f z_f g_z \bar{f} \gamma^{\mu} f Z_{\mu}^{\prime},
\ea
where $f=e_{R}^{j},l_{L}^{j},u_{R}^{j},d_{R}^{j},q_{L}^{j}$ and
$q_L^{j}=(u_{L}^{j},d_{L}^{j})\,\,,l_L^{j}=(\nu_{L}^{j},e_{L}^{j})$.
The coefficients $z_{u},z_{d}$ are the charges of the right-handed up and
down quarks, respectively, while the $z_q$ coefficients are the charges
of the left-handed quarks. $g_z$ is the $Z^{\prime}$ coupling constant.
We can write the Lagrangean for the $Z^{\prime}$-lepton-quark
interactions as follows
\ba
{\mathcal{L}}_{Z^{\prime}}=\sum_{j}g_z 
Z^{\prime}_{\mu}\left[z_{e_{R}^{j}} \bar{e}_{R}^{j}\gamma^{\mu}e_{R}^{j}+
z_{l_{L}^{j}}\bar{l}_{L}^{j}\gamma^{\mu}l_{L}^{j} +z_{u_{R}^{j}} 
\bar{u}_{R}^{j}\gamma^{\mu}u_{R}^{j}+ z_{d_{R}^{j}} 
\bar{d}_{R}^{j}\gamma^{\mu}d_{R}^{j}
+ z_{q_{L}^{j}}\bar{Q}_{L}^{j}\gamma^{\mu}Q_{L}^{j}\right],
\nonumber\\
\ea
with $j$ being the generation index. The low energy spectrum of the model, as
discussed above, is assumed to be the same for the other models that we
analyze in parallel. As shown in (\ref{table1})  the field content of the
model is effectively that of the Standard Model plus 1 additional Higgs
doublet. The extra scalars  $\phi$, and $\zeta_H,\bar{\zeta}_H$ and the
right handed components $N_H$ and $\bar{N}_H$ are assumed to decouple. In
this simplified framework, the structure of the vertex

\beqn
\begin{tabular}{|c|rrrr|}
\hline
\bf{Field}&$U(1)_Y$&$U(1)_{\zeta^\prime}$&$U(1)_{\zeta}$&$U(1)_{Z^\prime}$\\
\hline
${Q}^i$ & $\frac{1}{6}$ &$ \frac{1}{2}$& $-\frac{1}{2}$& $ \frac{3}{5}$ \\
${L}^i$ &$-\frac{1}{2}$ &$-\frac{3}{2}$& $-\frac{1}{2}$& $ \frac{1}{5}$ \\
${U}^i$ &$-\frac{2}{3}$ &$ \frac{1}{2}$& $ \frac{1}{2}$& $-\frac{2}{5}$ \\
${D}^i$ & $\frac{1}{3}$ &$-\frac{3}{2}$& $ \frac{1}{2}$& $-\frac{4}{5}$ \\
${E}^i$ & $      1    $ &$ \frac{1}{2}$& $ \frac{1}{2}$& $-\frac{2}{5}$ \\
${N}^i$ & $      0    $ &$ \frac{5}{2}$& $ \frac{1}{2}$& $     0      $ \\
$\phi^i$ &$      0    $ &$ 0          $& $      0     $& $     0      $ \\
$\phi^0$ &$      0    $ &$ 0          $& $      0     $& $     0      $ \\
$H^U$ & $ \frac{1}{2}$&$     -1     $& $    0         $& $-\frac{1}{5}$ \\
$H^D$ & $-\frac{1}{2}$&$     1      $& $    0         $& $ \frac{1}{5}$ \\
$N_H$ &   $0$         &$\frac{5}{2}$ & $ \frac{1}{2}  $& $     0      $ \\
${\bar N}_H$ &   $0$       &-$\frac{5}{2}$& $-\frac{1}{2}$&$   0      $ \\
$ {\zeta}_H$ &   $0$       & $      0    $& $        1   $& $  1      $ \\
${\bar\zeta}_H$&  $0$      & $      0    $& $       -1   $& $- 1      $ \\
\hline
\end{tabular}
\label{table1}
\eeqn

is the following
\ba
-\frac{ig}{4\cos{\theta_W}}\bar{\psi}_{i}\gamma^{\mu}(g_V^{Z,Z^{\prime}} 
+g_A^{Z,Z^{\prime}}\gamma^{5})\psi V_{\mu},
\ea
where $V_{\mu}$ denotes generically the vector boson. In the Standard 
Model (SM) 
\ba
&&v_{u}^{\gamma}=\frac{2}{3} \hspace{3cm} a_{u}^{\gamma}=0\nonumber\\
&&v_{d}^{\gamma}=-\frac{1}{3} \hspace{3cm} a_{d}^{\gamma}=0\nonumber\\
&&v_{u}^{Z}=1-\frac{8}{3}\sin^2\theta_W \hspace{1cm} a_{u}^{Z}=-1\nonumber\\
&&v_{d}^{Z}=-1+\frac{4}{3}\sin^2\theta_W \hspace{1cm} a_{d}^{Z}=1\,.
\ea
We need to generalize this formalism to the case of the $Z^{\prime}$.

Our starting point is the covariant derivative in a basis where the three 
electrically-neutral gauge bosons $W_{\mu}^{3},B_{Y}^{\mu},B_{z}^{\mu}$
are
\ba
\hat{D}_{\mu}=\left[\partial_{\mu} -i g \left( W_{\mu}^{1}T^{1} + 
W_{\mu}^{2}T^{2} + W_{\mu}^{3}T^{3} \right) -i\frac{g_{Y}}{2}\hat{Y} 
B_{Y}^{\mu}-i\frac{g_{z}}{2}\hat{z} B_{z}^{\mu} \right]
\ea
and we denote with $g,g_Y, g_z$ the couplings of $SU(2)$, $U(1)_Y$ and
$U(1)_z$, with  $\tan\theta_W=g_Y/g$. After the diagonalization of the
mass matrix we have
\ba
\left( \begin{array}{c}
A_{\mu} \\
Z_{\mu}  \\
Z^{\prime}_{\mu}
\end{array} \right)
=
\left( \begin{array}{ccc}
\sin\theta_W & \cos\theta_W & 0\\
\cos\theta_W & -\sin\theta_W & \varepsilon \\
-\varepsilon\sin\theta_W& \varepsilon\sin\theta_W & 1
\end{array} \right)
\left( \begin{array}{c}
W^{3}_{\mu} \\
B^{Y}_{\mu}  \\
B^{z}_{\mu}
\end{array} \right)
\ea
where $\varepsilon$ is defined as a perturbative parameter
\ba
&&\varepsilon=\frac{\delta M^2_{Z 
Z^{\prime}}}{M^2_{Z^{\prime}}-M^2_{Z}}\nonumber\\
&&M_Z^2=\frac{g^2}{4 
\cos^2\theta_W}(v_{H_1}^2+v_{H_2}^2)\left[1+O(\varepsilon^2)\right]
\nonumber\\
&&M_{Z^{\prime}}^2=\frac{g_z^2}{4}(z_{H_1}^2 
v_{H_1}^2+z_{H_2}^2v_{H_2}^2+z_{\phi}^2 
v_{\phi}^2)\left[1+O(\varepsilon^2)\right]
\nonumber\\
&&\delta M^2_{Z Z^{\prime}}=-\frac{g g_z}{4\cos\theta_W}(z_{H_1}^2 
v_{H_1}^2+z_{H_2}^2v_{H_2}^2).
\ea
Then we define
\ba
g=\frac{e}{\sin\theta_W} \hspace{1cm} g_Y=\frac{e}{\cos\theta_W},
\ea
and we construct the $W^{\pm}$ charge eigenstates and the corresponding 
generators $T^{\pm}$ as usual
\ba
&&W^{\pm}=\frac{W_1\mp iW_2}{\sqrt{2}}\nonumber\\
&&T^{\pm}=\frac{T_1\pm iT_2}{\sqrt{2}},
\ea
with the rotation matrix
\ba
\left( \begin{array}{c}
W^{3}_{\mu} \\
B^{Y}_{\mu}  \\
B^{z}_{\mu}
\end{array} \right)
=
\left( \begin{array}{ccc}
\frac{\sin\theta_W (1+\varepsilon^2)}{1+\varepsilon^2} & 
\frac{\cos\theta_W}{1+\varepsilon^2} & \varepsilon 
\frac{\cos\theta_W}{1+\varepsilon^2}\\
\frac{\cos\theta_W(1+\varepsilon^2)}{1+\varepsilon^2} & 
-\frac{\sin\theta_W}{1+\varepsilon^2} & \varepsilon 
\frac{\sin\theta_W}{1+\varepsilon^2}\\
0 & \frac{\varepsilon}{1+\varepsilon^2}& \frac{1}{1+\varepsilon^2}
\end{array} \right)
\left( \begin{array}{c}
A_{\mu} \\
Z_{\mu}  \\
Z^{\prime}_{\mu}
\end{array} \right)
\ea
from the interaction to the mass eigenstates. 
Substituting these expression in the covariant derivative we obtain
\ba
&&\hat{D}_{\mu}=\left[\partial_{\mu} -i A_{\mu} \left(g T_3\sin\theta_W+ 
g_Y\cos\theta_W \frac{\hat{Y}}{2}\right)
-ig\left(W^{-}_{\mu}T^{-}+ W^{+}_{\mu}T^{+}\right)\right.
\nonumber\\
&&\hspace{1cm}\left.-iZ_{\mu}\left( g\cos\theta_W T_{3} -g_Y \sin\theta_W 
\frac{\hat{Y}}{2}
+g_z \varepsilon\frac{\hat{z}}{2}\right)\right.
\nonumber\\
&&\hspace{1cm}\left.-iZ^{\prime}_{\mu}\left(-g\cos\theta_W 
T_{3}\varepsilon +g_Y\sin\theta_W \frac{\hat{Y}}{2}\varepsilon
+g_z\frac{\hat{z}}{2}\right)\right]
\ea
where we have neglected all the $O(\varepsilon^2)$ terms.
Sending $g_z\rightarrow 0$ and $\varepsilon\rightarrow 0$ 
we obtain the SM expression for the covariant derivative.
The next step is to separate left and right contributions in the 
interactions between the fermions and the $Z^{\prime}$ boson.
Hence for the quarks and the leptons we can write an interaction
Lagrangean of the type
\ba
&&{\mathcal{L}}_{int}=
\bar{Q}_{L}^{j}N^{Z}_{L}\gamma^{\mu}Q^{j}_{L} Z_{\mu}
+\bar{Q}_{L}^{j}N^{Z^{\prime}}_{L}\gamma^{\mu}Q^{j}_{L} Z^{\prime}_{\mu}
+\bar{u}_{R}^{j}N^{Z}_{u,R}\gamma^{\mu}u^{j}_{R} Z_{\mu}
\nonumber\\
&&\hspace{1cm}
+\bar{d}_{R}^{j}N^{Z}_{d,R}\gamma^{\mu}d^{j}_{R} Z_{\mu}
+\bar{u}_{R}^{j}N^{Z^{\prime}}_{u,R}\gamma^{\mu}u^{j}_{R} Z^{\prime}_{\mu}
+\bar{d}_{R}^{j}N^{Z^{\prime}}_{d,R}\gamma^{\mu}d^{j}_{R} Z^{\prime}_{\mu}
\nonumber\\
&&\hspace{1cm}
+\bar{Q}_{L}^{j}N^{\gamma}_{L}\gamma^{\mu}Q^{j}_{L} A_{\mu}
+\bar{u}_{R}^{j}N^{\gamma}_{u,R}\gamma^{\mu}u^{j}_{R} A_{\mu}
+\bar{d}_{R}^{j}N^{\gamma}_{d,R}\gamma^{\mu}d^{j}_{R} A_{\mu}
\nonumber\\
&&\hspace{1cm}
+\bar{l}_{L}^{j}N^{\gamma}_{L}\gamma^{\mu}l^{j}_{L} A_{\mu}
+\bar{e}_{R}^{j}N^{\gamma}_{e,R}\gamma^{\mu}e^{j}_{R} A_{\mu}
\nonumber\\
&&\hspace{1cm}
+\bar{l}_{L}^{j}N^{Z}_{L,lep}\gamma^{\mu}l^{j}_{L} Z_{\mu}
+\bar{l}_{L}^{j}N^{Z^{\prime}}_{L,lep}\gamma^{\mu}l^{j}_{L} Z^{\prime}_{\mu}
\nonumber\\
&&\hspace{1cm}
+\bar{e}_{R}^{j}N^{Z}_{e,R}\gamma^{\mu}e^{j}_{R} Z_{\mu}
+\bar{e}_{R}^{j}N^{Z^{\prime}}_{e,R}\gamma^{\mu}e^{j}_{R} Z^{\prime}_{\mu}
\ea
\begin{figure}[t]
{\includegraphics[%
  width=10cm,
  angle=-90]{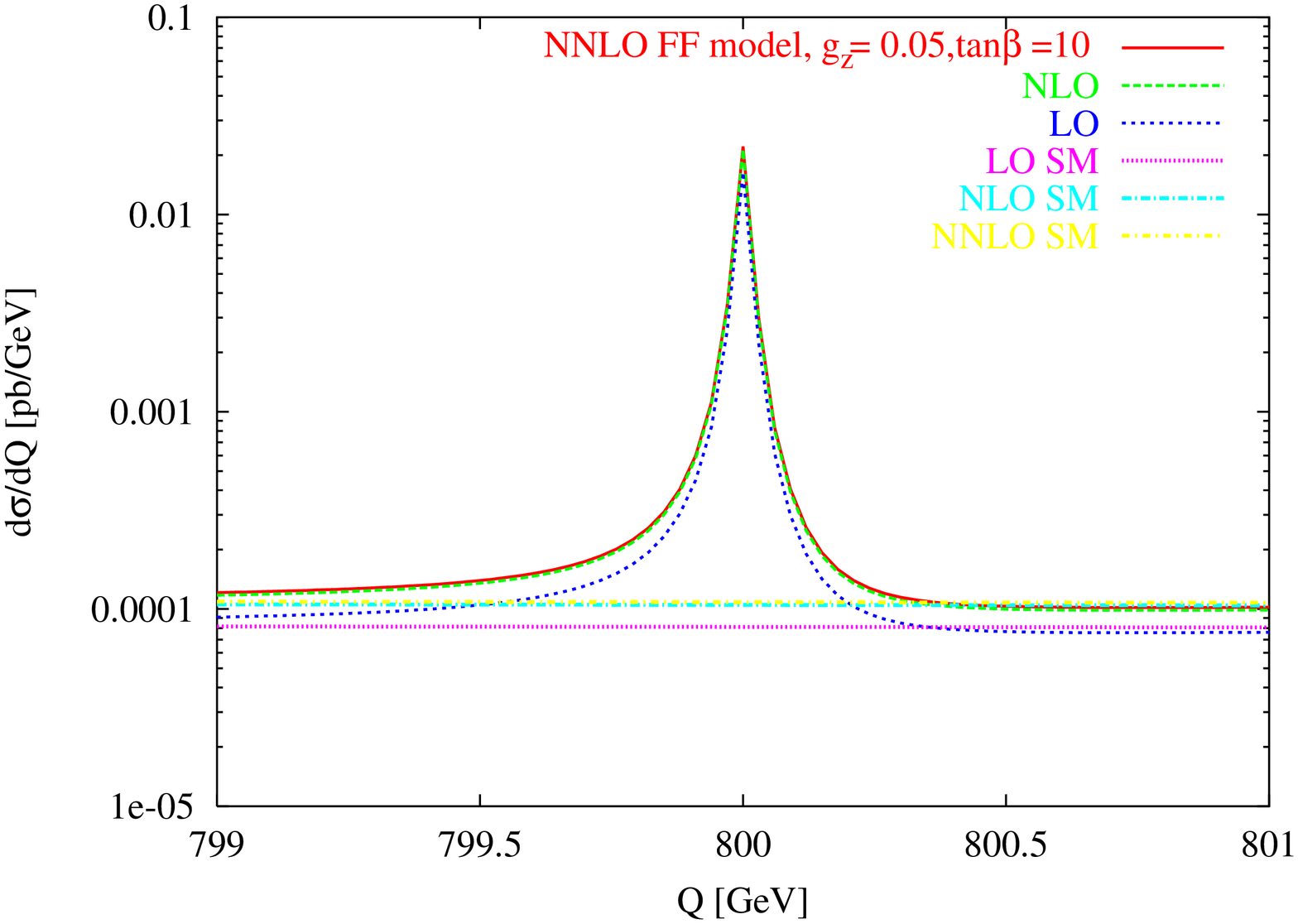}}
\caption{Plot of the LO, NLO and NNLO cross section for the free 
fermionic model with $M_{Z^{\prime}}=800$ GeV.}
\label{cross01}
\end{figure}
where for the quarks we have
\ba
&&N^{Z,j}_{L}=-i\left( g\cos\theta_W T^{L}_{3} -g_Y \sin\theta_W 
\frac{\hat{Y}^{L}}{2} + g_z \varepsilon\frac{\hat{z}^{L}}{2}\right)
\nonumber\\
&&N^{Z^{\prime},j}_{L}=-i\left(-g\cos\theta_W T^{L}_{3}\varepsilon
+g_Y\sin\theta_W  
\frac{\hat{Y}^{L}}{2}\varepsilon+g_z\frac{\hat{z}^{L}}{2}\right)
\nonumber\\
&&N^{Z}_{u,R}=-i\left(
-g_Y \sin\theta_W \frac{\hat{Y}^{u,R}}{2}+g_z 
\varepsilon\frac{\hat{z}^{u,R}}{2}\right)
\nonumber\\
&&N^{Z}_{d,R}=-i\left(
-g_Y \sin\theta_W \frac{\hat{Y}^{d,R}}{2}+g_z
\varepsilon\frac{\hat{z}^{d,R}}{2}\right),\,
\ea
and similar expressions for the leptons. We rewrite the vector and the 
axial coupling of the $Z$ and $Z^{\prime}$ bosons to the quarks as
\ba
&&\frac{-i g}{4 c_w}\gamma^{\mu} {g_V}^{Z,j}=\frac{-i g}{c_w}
\frac{1}{2}\left[c_w^2 
T_3^{L,j}-s_w^2(\frac{\hat{Y}^{j}_L}{2}+\frac{\hat{Y}^{j}_R}{2})
+\varepsilon \frac{g_z}{g} c_w 
(\frac{\hat{z}_{L,j}}{2}+\frac{\hat{z}_{R,j}}{2})\right]\gamma^{\mu}
\nonumber\\
&&\frac{-i g}{4 c_w}\gamma^{\mu}\gamma^{5} {g_A}^{Z,j}=\frac{-i g}
{c_w}\frac{1}{2}\left[-c_w^2 T_3^{L,j}
-s_w^2(\frac{\hat{Y}^{j}_R}{2}-\frac{\hat{Y}^{j}_L}{2})
+\varepsilon \frac{g_z}{g} c_w 
(\frac{\hat{z}_{R,j}}{2}-\frac{\hat{z}_{L,j}}{2})\right]\gamma^{\mu}\gamma^{5}
\nonumber\\
&&\frac{-i g}{4 c_w}\gamma^{\mu} {g_V}^{Z^{\prime},j}=\frac{-i g}{c_w}
\frac{1}{2}\left[ -\varepsilon c_w^2 T_3^{L,j}
+\varepsilon s_w^2(\frac{\hat{Y}^{j}_L}{2}+\frac{\hat{Y}^{j}_R}{2})
+\frac{g_z}{g}c_w(\frac{\hat{z}_{L,j}}{2}+
\frac{\hat{z}_{R,j}}{2})\right]\gamma^{\mu}
\nonumber\\
&&\frac{-i g}{4 c_w}\gamma^{\mu}\gamma^{5} {g_A}^{Z^{\prime},j}=\frac{-i 
g}{c_w} \frac{1}{2}\left[ \varepsilon c_w^2 T_3^{L,j}
+\varepsilon s_w^2(\frac{\hat{Y}^{j}_R}{2}-\frac{\hat{Y}^{j}_L}{2})
+\frac{g_z}{g}c_w(\frac{\hat{z}_{R,j}}{2}-
\frac{\hat{z}_{L,j}}{2})\right]\gamma^{\mu}\gamma^{5},
\nonumber\\
\ea
where $j$ is an index which represents the quark or the lepton and we have 
set $\sin\theta_W=s_w,\cos\theta_W=c_w$ for brevity.

The decay rates into leptons for the $Z$ and the $Z^{\prime}$ are
universal and are given by
\ba
&&\Gamma({\cal Z}\rightarrow l\bar{l})=\frac{g^2}{192\pi c_w^2}
M_{{\cal Z}}\left[(g_{V}^{{\cal Z},l})^2+(g_{A}^{{\cal Z},l})^2\right]=
\frac{\alpha_{em}}{48 s_w^2 c_w^2}M_{{\cal Z}}\left[(g_{V}^{{\cal Z},l})^2+
(g_{A}^{{\cal Z},l})^2\right]\,,
\nonumber\\
&&\Gamma({\cal Z}\rightarrow \psi_i\bar{\psi_i})=\frac{N_c\alpha_{em}}{48 
s_w^2 c_w^2}
M_{{\cal Z}}\left[(g_{V}^{{\cal Z},\psi_i})^2+(g_{A}^{{\cal 
Z},\psi_i})^2\right]\times\nonumber\\
&&\hspace{3cm}\left[1+ \frac{\alpha_s(M_{{\cal Z}})}{\pi}
+1.409\frac{\alpha_s^2(M_{{\cal 
Z}})}{\pi^2}-12.77\frac{\alpha_s^3(M_{\cal Z})}{\pi^3}\right],\,
\ea
where $i=u,d,c,s$ and ${\cal Z}=Z,Z^{\prime}$.

For the $Z^{\prime}$ and $Z$ decays into heavy quarks we obtain
\ba
&&\Gamma({\cal Z}\rightarrow b\bar{b})=\frac{N_c\alpha_{em}}{48 s_w^2 c_w^2}
M_{{\cal Z}}\left[(g_{V}^{{\cal Z},b})^2+(g_{A}^{{\cal 
Z},b})^2\right]\times\nonumber\\
&&\hspace{3cm}\left[1+ \frac{\alpha_s(M_{{\cal 
Z}})}{\pi}+1.409\frac{\alpha_s^2(M_{{\cal Z}})}{\pi^2}
-12.77\frac{\alpha_s^3(M_{\cal Z})}{\pi^3}\right]\,,
\nonumber\\
&&\Gamma({\cal Z}\rightarrow t\bar{t})=\frac{N_c\alpha_{em}}{48 s_w^2 c_w^2}
M_{{\cal Z}}\sqrt{1 - 4 \frac{m_t^2}{M_{{\cal Z}}^{2}} }\times\nonumber\\
&&\hspace{3cm}\left[(g_{V}^{{\cal Z},t})^2\left(1 + 2 
\frac{m_t^2}{M_{{\cal Z}}^{2}}\right)
+(g_{A}^{{\cal Z},t})^2\left(1 - 4 \frac{m_t^2}{M_{{\cal 
Z}}^{2}}\right)\right]\times\nonumber\\
&&\hspace{3cm}\left[1+ \frac{\alpha_s(M_{{\cal 
Z}})}{\pi}+1.409\frac{\alpha_s^2(M_{{\cal Z}})}{\pi^2}
-12.77\frac{\alpha_s^3(M_{\cal Z})}{\pi^3}\right]\,.\nonumber\\
\ea
The total hadronic widths are defined by
\ba
\Gamma_Z\equiv\Gamma(Z\rightarrow hadrons)=\sum_{i}\Gamma(Z\rightarrow \psi_i\bar{\psi_i})
\nonumber\\
\Gamma_{Z^\prime}\equiv\Gamma(Z^{\prime}\rightarrow 
hadrons)=\sum_{i}\Gamma(Z^{\prime}\rightarrow \psi_i\bar{\psi_i})
\ea
where we refer to hadrons not containing bottom and top quarks (i.e. 
$i=u,d,c,s$). We also ignore electroweak corrections and all fermion 
masses with the exception
of the top-quark mass, while we have included the relevant QCD corrections.
Similarly to \cite{Carena} we have considered only tree level
decays into fermions, assuming that the decays into particles
other than the SM fermions are either invisible or are negligible
in their branching ratios, then the total decay rate for the $Z$ and $Z^{\prime}$ is given by
\begin{figure}[t]
{\includegraphics[%
  width=9cm,
  angle=-90]{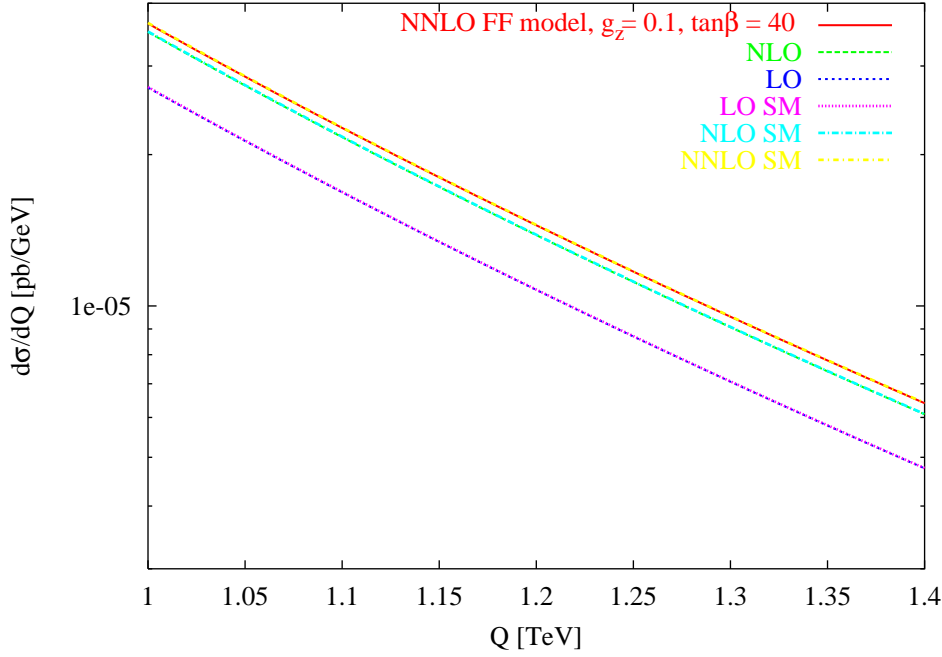}}
\caption{Plot of the LO, NLO and NNLO cross section for the free 
fermionic model with $M_{Z^{\prime}}=800$ GeV in the TeVs region.}
\label{cross02}
\end{figure}
\ba
&&\Gamma_Z=\sum_{i=u,d,c,s}\Gamma(Z\rightarrow 
\psi_i\bar{\psi_i})+\Gamma(Z\rightarrow b\bar{b})
+3\Gamma(Z\rightarrow l\bar{l})+3\Gamma(Z\rightarrow \nu_l\bar{\nu_l})
\nonumber\\
&&\Gamma_{Z^{\prime}}=\sum_{i=u,d,c,s}\Gamma(Z^{\prime}\rightarrow 
\psi_i\bar{\psi_i})+\Gamma(Z^{\prime}\rightarrow b\bar{b})
+\Gamma(Z^{\prime}\rightarrow t\bar{t})+3\Gamma(Z^{\prime}\rightarrow l\bar{l})
+3\Gamma(Z^{\prime}\rightarrow \nu_l\bar{\nu_l}).
\nonumber\\
\ea

\begin{figure}[t]
{\includegraphics[%
  width=10cm,
  angle=-90]{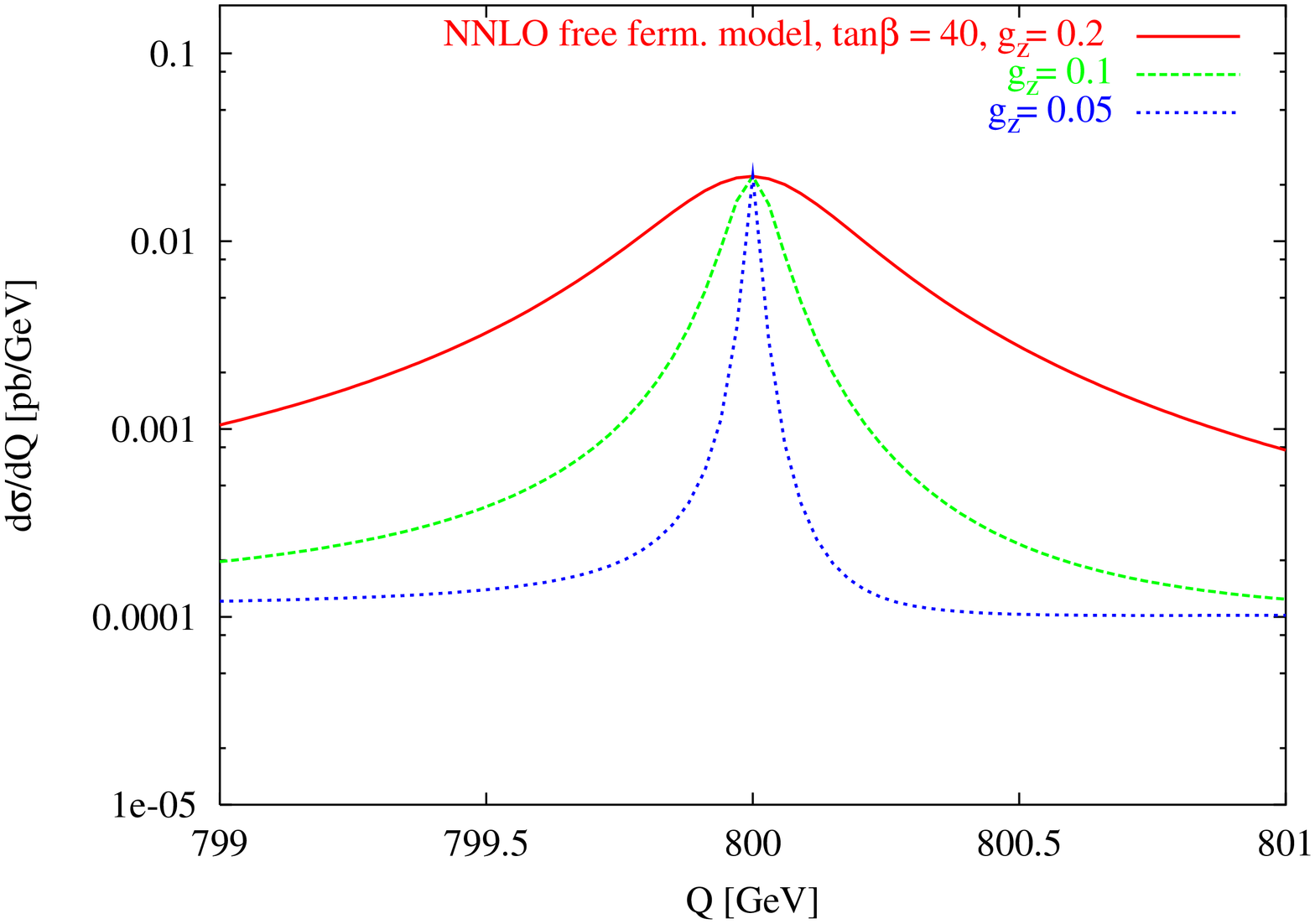}}
\caption{Free fermionic model at the LHC, $\tan{\beta}=40$}
\label{cross03}
\end{figure}
We also recall that the point-like cross sections for the photon, the SM  
$Z_0$ and the new $Z^\prime$  gauge boson are written as
\ba
&&\sigma_{\gamma}(Q^2)=\frac{4\pi\alpha_{em}^2}{3 Q^4}\frac{1}{N_c}
\nonumber\\
&&\sigma_{Z}(Q^2,M_Z^2)=\frac{\pi\alpha_{em}}{4 M_{Z}\sin^2\theta_W 
\cos^2\theta_W N_c}
\frac{\Gamma_{Z\rightarrow \bar{l} l}}{(Q^2-M_Z^2)^2 + M_Z^2 \Gamma_Z^2}
\nonumber\\
&&\sigma_{Z,\gamma}(Q^2,M_Z^2)=\frac{\pi\alpha_{em}^2}{6} 
\frac{(1-4\sin^2{\theta_W})}{\sin^2{\theta_W}\cos^2{\theta_W}}
\frac{(Q^2-M_Z^2)}{N_C Q^2(Q^2-M_Z^2)^2+M_Z^2\Gamma_Z^2},\nonumber\\
\ea
 where $N_C$ is the number of colours, and 
 \ba
&&\sigma_{{Z^{\prime}}}(Q^2)=\frac{\pi\alpha_{em}}{4 
M_{{Z^{\prime}}}\sin^2\theta_W \cos^2\theta_W N_c}
\frac{\Gamma_{{Z^{\prime}}\rightarrow \bar{l} 
l}}{(Q^2-M_{Z^{\prime}}^2)^2 + M_{Z^{\prime}}^2 \Gamma_{Z^{\prime}}^2}
\nonumber\\
&&\sigma_{{Z^{\prime}},\gamma}(Q^2)=\frac{\pi\alpha_{em}^2}{6 N_c} 
\frac{g_V^{Z^{\prime},l}g_V^{\gamma,l}}{\sin^2{\theta_W}\cos^2{\theta_W}}
\frac{(Q^2-M_{Z^{\prime}}^2)}{Q^2(Q^2-M_{Z^{\prime}}^2)^2+
M_{Z^{\prime}}^2\Gamma_{Z^{\prime}}^2},\nonumber\\
&&\sigma_{{Z^{\prime}},Z}(Q^2)=\frac{\pi\alpha_{em}^2}{96}
\frac{\left[g_V^{Z^{\prime},l}g_V^{Z,l}+
g_A^{Z^{\prime},l}g_A^{Z,l}\right]}{\sin^4{\theta_W}\cos^4{\theta_W}N_c}
\frac{(Q^2-M_Z^2)(Q^2-M^2_{Z^{\prime}}) 
+M_Z\Gamma_{Z}M_{Z^{\prime}}\Gamma_{Z^{\prime}}}
{\left[(Q^2-M_{Z^{\prime}}^2)^2 + M_{Z^{\prime}}^2\Gamma_{Z^{\prime}}^2\right]
\left[(Q^2-M_Z^2)^2 + M_Z^2\Gamma_{Z}^2\right]}.\nonumber\\
\ea
The contributions such as $Z,\gamma$ and similar denote the interference terms.
At LO (or leading order) the process proceeds through the $q \bar{q}$ annihilation channel and is
$O(1)$ in the strong coupling constant $\alpha_s$. The NLO (or next-to-leading order) corrections
involve virtual corrections with one gluon exchanged in the 
initial state and real emissions involving a single gluon, which 
is integrated over phase space.  These corrections are $O(\alpha_s)$ in the strong coupling. The
change induced by moving from LO to NLO amounts to approximately 
a 20 to 30 \% in the numerical value of the cross section that 
we consider. At the highest accuracy, we use in our analysis  partonic contributions
with hard scattering computed at NNLO, or $O(\alpha_s^2)$.
At this order typical real emissions involve 2 partons in 
the final state - which are integrated over their phase space- and 
two-loop virtual corrections at the same perturbative order. 
The cross section for the invariant mass distributions factorizes at a
perturbative level in terms of a NNLO (next-to-next-to-leading, or  $O(\alpha_s^2)$) contribution
$W_V$ (which takes into account all the
initial state emissions of real gluons and all the virtual corrections)
and a point-like cross section. The computation of $W_V$ can be found in \cite{Van_Neerven1}
to which we refer for more details. A similar factorization holds
also for the total cross section if we use the narrow width approximation. 
At NLO (next-to-leading order, or $O(\alpha_s)$).
The colour-averaged inclusive differential cross section
for the reaction $p +p \rightarrow l_1 +l_2 +X $, is given by
\ba
\frac{d\sigma}{dQ^2}=\tau \sigma_{V}(Q^2,M_V^2)
W_{V}(\tau,Q^2)\hspace{1cm} \tau=\frac{Q^2}{S},
\ea
where all the hadronic initial state information is contained in the
hadronic structure function which is defined as
\ba
W_{V}(\tau,Q^2)=\sum_{i,j} \int_{0}^{1}dx_1 \int_0^1 dx_2 \int_{0}^{1}dx
\delta(\tau-x x_1 x_2)
PD_{i,j}^{V}(x_1,x_2,\mu_F^2)\Delta_{i,j}(x,Q^2,\mu_F^2)\,,
\nonumber\\
\ea
where the quantity $PD_{i,j}^{V}(x_1,x_2,\mu_F^2)$ contains all the
information about the parton distribution functions and their evolution up to
the $\mu_F^2$ scale, while the functions $\Delta_{i,j}(x,Q^2,\mu_F^2)$ are
the hard scatterings. This factorization formula is universal for invariant
mass distributions mediated by s-channel exchanges of neutral or charged
currents. The hard scatterings can be expanded in a series in terms of the
running coupling constant $\alpha_s(\mu_R^2)$ as
\ba
\Delta_{i,j}(x,Q^2,\mu_F^2)=\sum_{n=0}^{\infty}
\alpha_s^n(\mu_R^2)\Delta^{(n)}_{i,j}(x,Q^2,\mu_F,\mu_R^2)\,.
\ea
In principle, factorization and renormalization scales should be kept
separate in order to determine the overall scale dependence of the
results. However, as we are going to show, the high-end of the Drell-Yan
distribution is not so sensitive to these higher order corrections, at
least for the models that we have studied.
\section{Numerical Results}
In our analysis we have decided to compare our results with a series of
models introduced in \cite{Carena}. We refer to this work for more
details concerning their general origin. We just mention that the
construction of  models with extra $Z^\prime$ using a
bottom-up approach is, in general, rather
straightforward, being based mostly on the principle of cancellation of
the gauge cubic $U(1)_{Z^\prime}^3$ and mixed  anomalies. One of the most
economical ways to proceed is to introduce just one additional $SU(2)_W$
Higgs doublet and an extra scalar (weak) singlet, as in
\cite{Appelquist}, and one right-handed neutrino per generation in order
to generate reasonable operators for their Majorana and Dirac masses.
However, more general solutions of the anomaly equations are possible
by  enlarging the fermion spectrum and/or enlarging the scalar sector \cite{group1}.
In \cite{Carena} the scalar sector is enlarged with 2 Higgs doublets and
one (weak) scalar singlet.

Anomalous constructions, instead, require a
different approach and several phenomenological analysis have been
presented recently \cite{CIM1,CIM2,ACG,CGM} that try to identify the
signature of these peculiar realizations. In the anomalous models, due
to the absence of the non-resonant behaviour of the s-channel
(at least in the double prompt photon production), the chiral
anomaly induces a unitarity growth which should be present in correlated
studies of other channels \cite{CGM}. For non anomalous  $Z^\prime$ the
phenomenological predictions are, as we are going to show, rather similar
for all the models - at least in the mass invariant distributions in Drell-Yan -
and the possibility to identify the underlying
interaction requires a careful study of the forward-backward and/or
charge asymmetries \cite{Petriello:2008zr}.
This is not going to be an easy task at the LHC,
given the size of the cross section at the tail of the invariant mass
distribution, the rather narrow widths, and given the presence of both
theoretical and experimental errors in the parton distributions (pdf's),
unless the gauge coupling is quite sizeable ($O(1)$). We refer to
\cite{CCG1} for an accurate analysis of the experimental errors on the
pdf's in the case of the Z peak. It has been shown that the errors on the
pdf's are comparable with the overall reduction of the cross section as
we move from the NLO to the NNLO.

These source of ambiguities, known as experimental errors, unfortunately
do not take into consideration the theoretical errors due to the
implementation of the solution of the DGLAP in the evolution codes, which
amount to a theoretical uncertainty  \cite{CCG2}. Once all these sources
of indeterminations are combined together, the expected error on the Z
peak is likey to be much larger than 3 $\%$. Given the large amount of
data that will accumulate in the first runs (for $Q=M_Z$), which will soon reduce the
statistical errors on the measurements far below the 0.1 $\%$ value, there
will be severe issues to be addressed also from the theoretical side in
order to match this far larger experimental accuracy.  The possibility to
use determinations of the pdf's on the Z peak for further studies of the $Z^\prime$ resonances at
larger invariant mass values of the lepton pair, have to face several
additional issues, such as the presence of an additional scale, which
is $Q=M_{Z^\prime}$, new respect to the $Q=M_Z$ scale used as a
benchmark for partonometry in the first accelerator runs. We remind that
logarithms of these two scales may also play a role especially if
$M_{Z^\prime}$ is far larger than $M_Z$. With these words of caution in
mind we proceed with our exploration of the class of models that we have
selected, starting from the string model and then analizing the bottom-up
models mentioned above \cite{Carena}. These are studied in the limit
$z_{H_1}=z_{H_2}=0$, with the mass of the extra $Z^\prime$ generated only
by the extra singlet scalar $\phi$. In the string model, as one can see
from  (\ref{table1}), only the two Higgses $H_U$ and $H_D$ contribute to
the mass of the new gauge boson. The differences between these two types
of models are, however, not relevant for this analysis, since the mass of
the extra gauge boson is essentially a free parameter in both cases.
\begin{figure}[t]
{\includegraphics[%
  width=9cm,
  angle=-90]{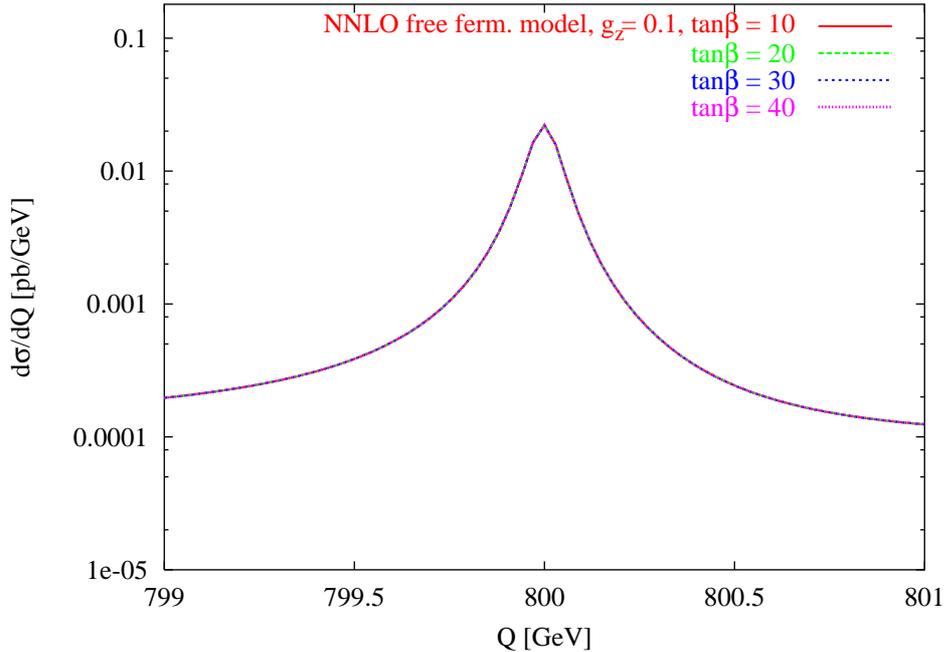}}
\caption{Free fermionic model at the LHC, $g_z=0.1$}
\label{cross04}
\end{figure}

The set of pdf's that we have used for our analysis is MRST2001 \cite{MRST}, which is
given in parametric form, evolved with CANDIA (see \cite{candia}). The models analyzed
numerically are the free fermionic one, ``F'', discussed in the previous
sections, and the ``$B-L$'', ``q +u'' , ``$10 + \bar{5}$'' and ``d-u'', using the
notations of  \cite{Carena}. 

Our results are organized in a series of
plots on the various resonances and in some tables which are useful in 
order to pin down the actual numerical value of the various cross 
sections at a given invariant mass. 

\subsection{$M_{Z^\prime}=0.8$ TeV}
We show in Fig. \ref{cross01} a plot of the $Z^\prime$ resonance around a 
typical value of 800 GeV for the $FF$ model and the SM. The coupling of 
the extra neutral gauge boson is taken to be $0.05$, with $\tan\beta=10$. 
We remark that the dependence of the resonance on this
second parameter is negligible. In fact the relevant parameters are the
coupling constant $g_{Z}$ and the mass $M_{Z^\prime}$. Notice that the 
width is very narrow ($\approx 1$ GeV) and basically invisible in an 
experimental analysis. Neverthless it is, at least theoretically, 
useful to try to characterize the signal and the background even in
this (and other similar) not favourable cases.
 
Assuming an integrated luminosity of $100 fb^{-1}/y$ after the first 3 
years at the LHC (per experiment), we would expect 10 background events
versus a signal of approximately 30 events. Notice that LO, NLO and
NNLO determinations are, essentially, coincident for all the practical
purposes. 

\begin{figure}
{\includegraphics[%
  width=10cm,
  angle=-90]{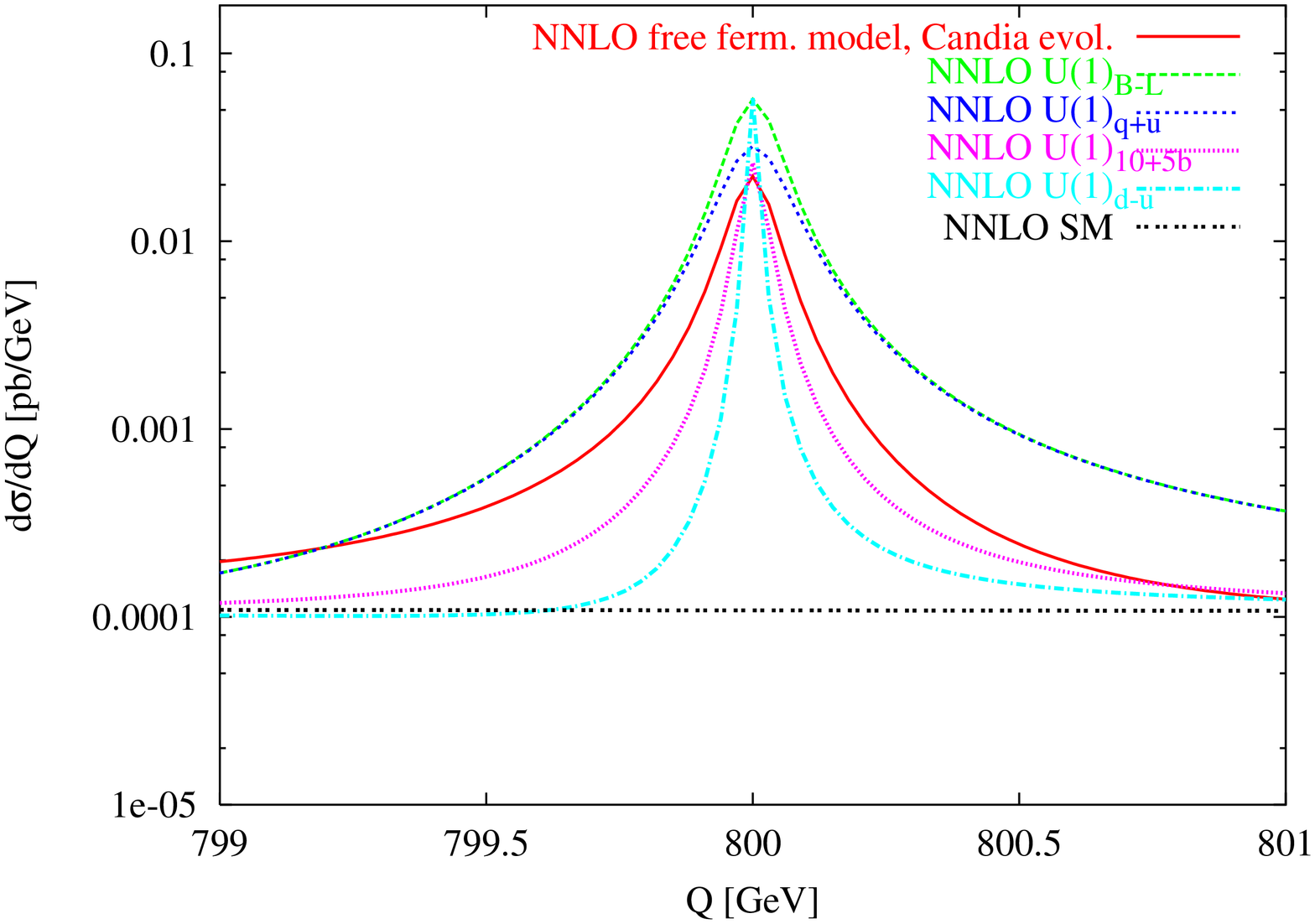}}
\caption{Free fermionic model at the LHC, $\tan{\beta}=40$ and $g_z=0.1$}
\label{cross06}
\end{figure}

\begin{figure}
{\includegraphics[%
  width=10cm,
  angle=-90]{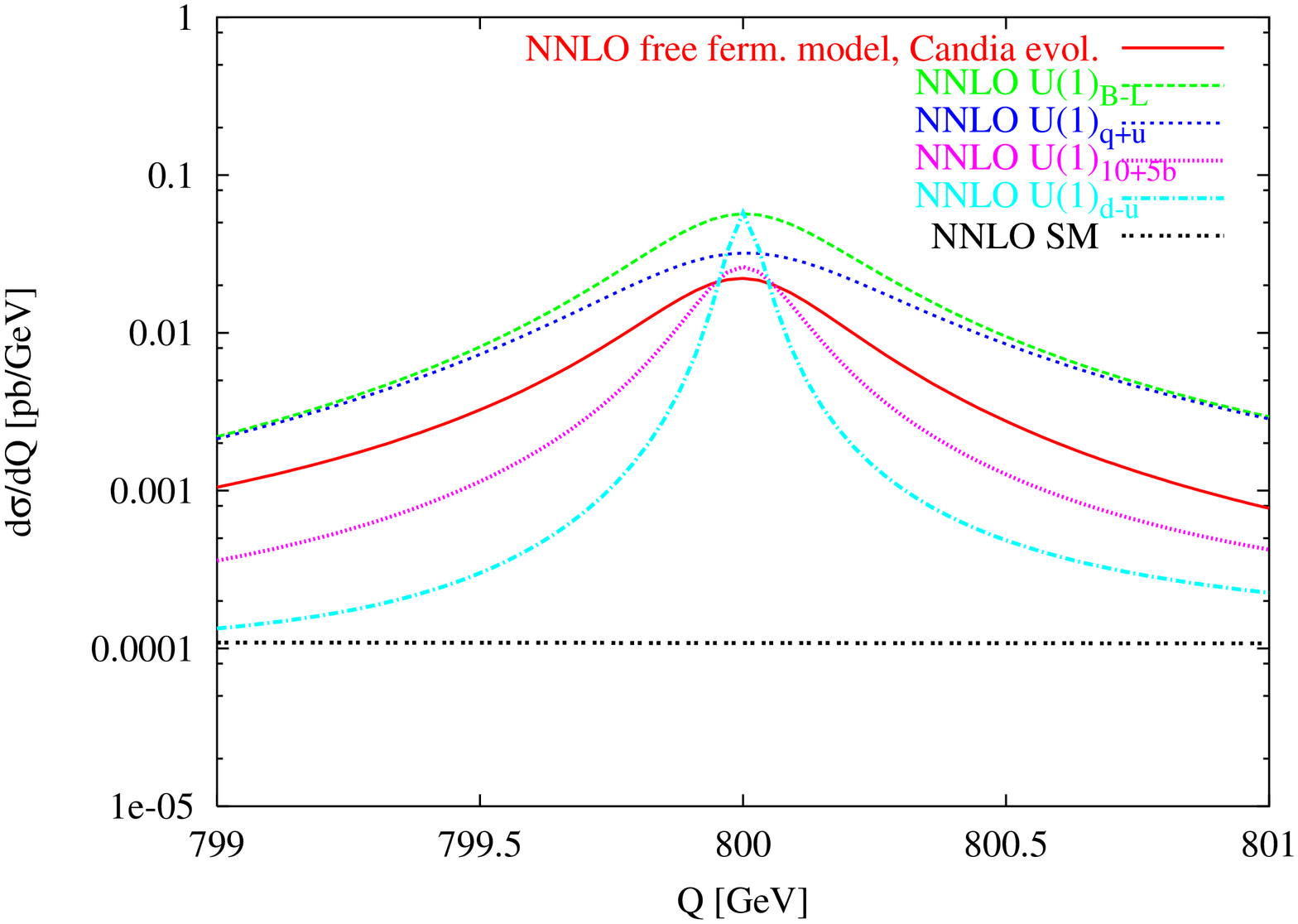}}
\caption{Free fermionic model at the LHC, $\tan{\beta}=40$ and $g_z=0.2$}
\label{cross07}
\end{figure}

In Fig. \ref{cross02} we show the tail of the distribution 
for a run with $M_{Z^\prime}=800$ GeV, where we have just modified
$\tan\beta$ and we have increased the coupling to $g_Z=0.1$. For $Q$
around $1.2$ TeV the determinations of the cross section in the FF and SM
models are basically overlapping as we move from LO to NLO and NNLO. The
LO determination in the SM moves up toward the FF result as we increase
the perturbative order. Also in this case, given the small size of the
cross section ($\approx 10^{-2}$ fb) the possibility to resolve these 
differences experimentally is remote.
\begin{figure}[t]
{\includegraphics[%
  width=10cm,
  angle=-90]{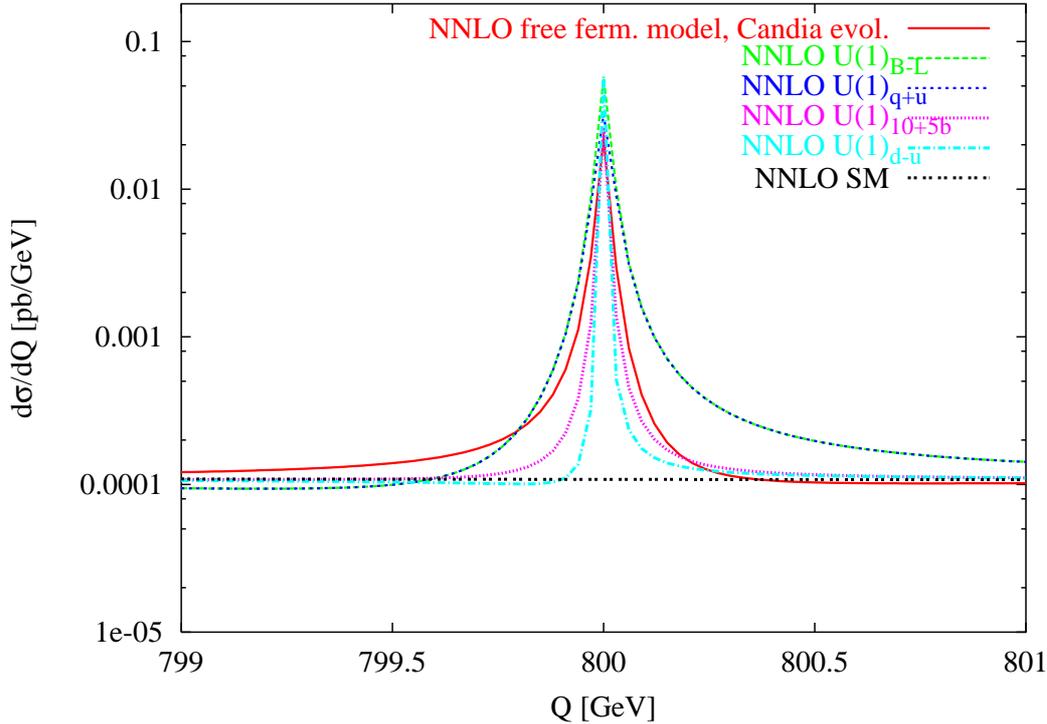}}
\caption{Free fermionic model at the LHC, $\tan{\beta}=40$ and $g_z=0.05$}
\label{cross05}
\end{figure}
In Fig. \ref{cross03} we vary the coupling constants of the extra $U(1)$
from a very small value $g_Z=0.05$ up to $g_Z=0.2$. The only variation in
the result is due to the width that increases from 1 to approximately 3-4
GeV's. Here we have chosen $\tan\beta=40$, and, as shown in Fig.
\ref{cross04} there is essentially no variation on the shape of the
resonance due to this variable. In Figs. \ref{cross05} and \ref{cross06} 
we perform a comparative study of  all the models and the SM background 
for a resonance mass of 800 GeV. There are only minor differences between 
the 4 bottom-up models and the FF model. The FF model shows a resonance 
curve which sits in the middle of all the determinations but is, for the 
rest, overlapping with the other curves. The ``$B-L$'' model, in all the
cases, shows a wider width among all, with the ``$q+u$'' model quite
similar to it. The ``$d-u$'' model has the narrowest width. This feature is
particularly obvious from Fig. \ref{cross07} where the result is
numerically smoothed out by the increased value of the coupling, which is
now doubled compared to Fig. \ref{cross06}.

\subsection{$M_{Z^{\prime}}= 1.2$ TeV}
\begin{figure}[t]
{\includegraphics[%
  width=10cm,
  angle=-90]{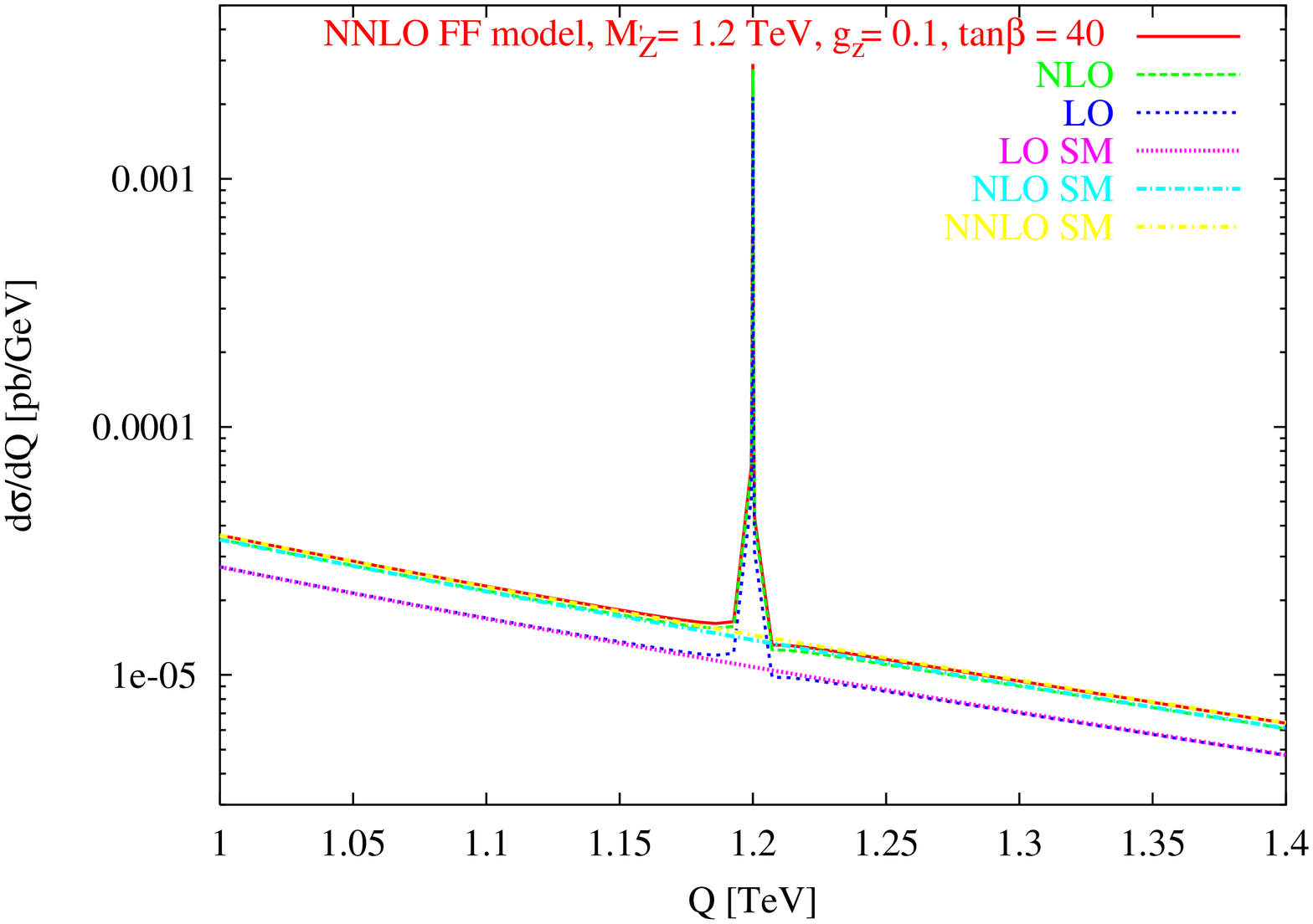}}
\caption{Free fermionic model at the LHC, $\tan{\beta}=40$ and $g_z=0.1$}
\label{cross08}
\end{figure}

We illustrate in the next 3 figures our results for the various models 
for $M_{Z^\prime}=1.2$ TeV. Fig. \ref{cross08} shows the behaviour of the 
cross section for this new mass value with $g_Z=0.1$, and the 
corresponding result for the SM case. The QCD corrections are very small 
and it is likely that the only role of these corrections, at these large 
$Q$ values, is to stabilize the dependence of the perturbative series 
from the factorization/renormalization scales. In our case we have 
chosen, for simplicity $\mu_F=\mu_R=Q$, where 
$\mu_R$ and $\mu_F$ are the renormalization and factorization scale, 
respectively. The separation of this dependence can be done as in 
\cite{CCG2}, by relating the coupling constants  at the two scales ($\mu_F,\mu_R$).

\begin{figure}[h]
{\includegraphics[%
  width=10cm,
  angle=-90]{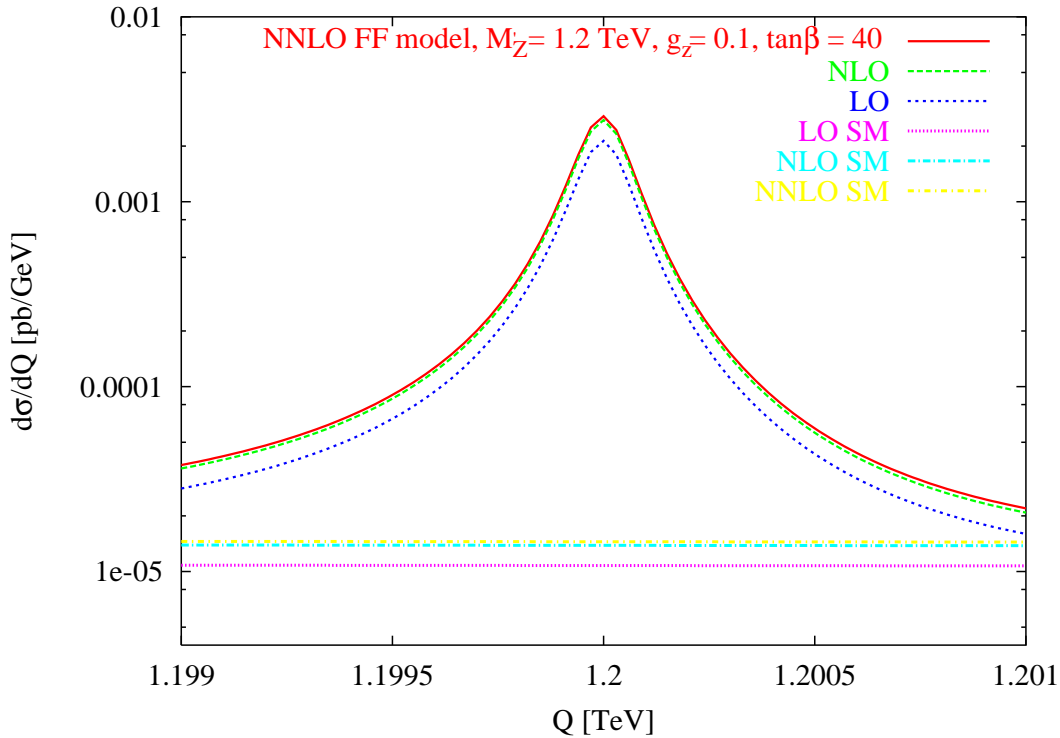}}
\caption{Free fermionic model at the LHC, $\tan{\beta}=40$ and $g_z=0.1$.
Shown are also the SM results through the same perturbative orders.}
\label{cross09}
\end{figure}
\begin{figure}[h]
{\includegraphics[%
  width=10cm,
  angle=-90]{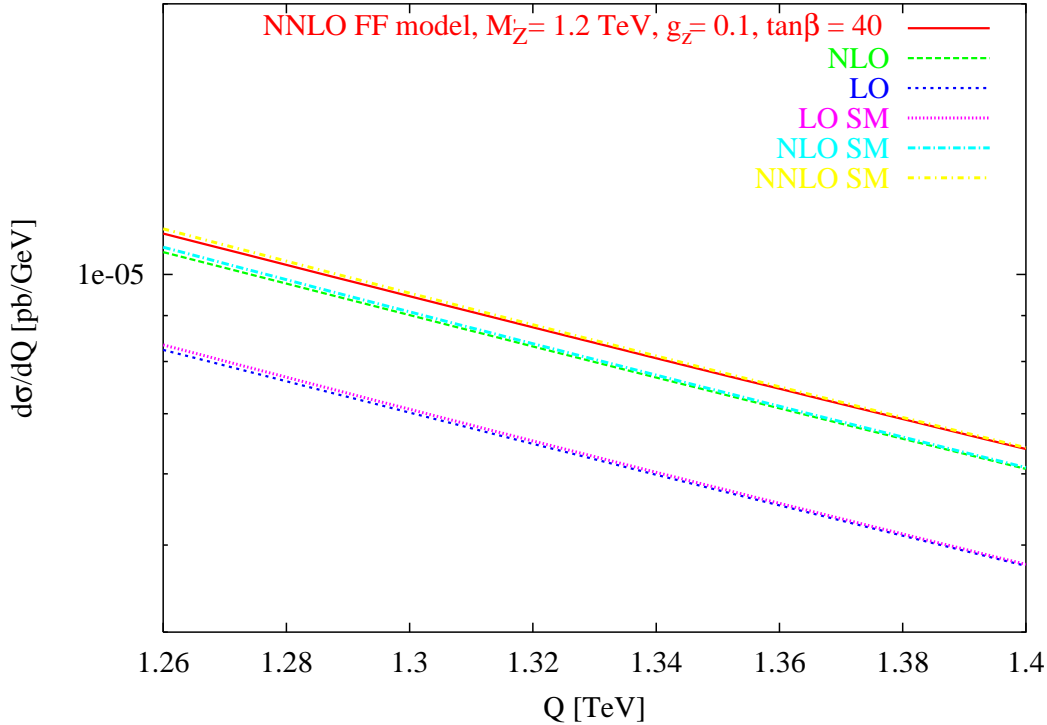}}
\caption{Free fermionic model at the LHC, $\tan{\beta}=40$ and $g_z=0.1$
and the corresponding SM results. The plot is a zooming of the resonance
shape shown in Fig. \ref{cross08}} \label{cross010}
\end{figure}
\begin{figure}[h]
{\includegraphics[%
  width=10cm,
  angle=-90]{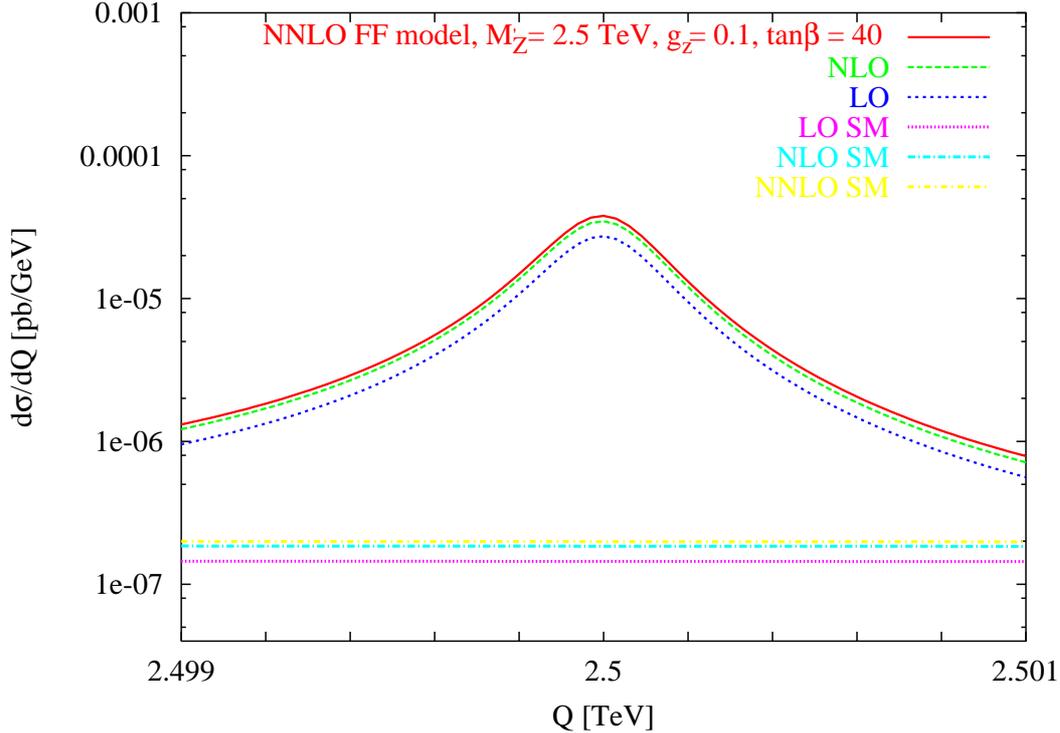}}
\caption{Free fermionic model and the corresponding SM results at all the
three orders for $M_{Z^\prime}=2.5$ TeV.}
\label{cross011}
\end{figure}
\begin{figure}[h]
{\includegraphics[%
  width=10cm,
  angle=-90]{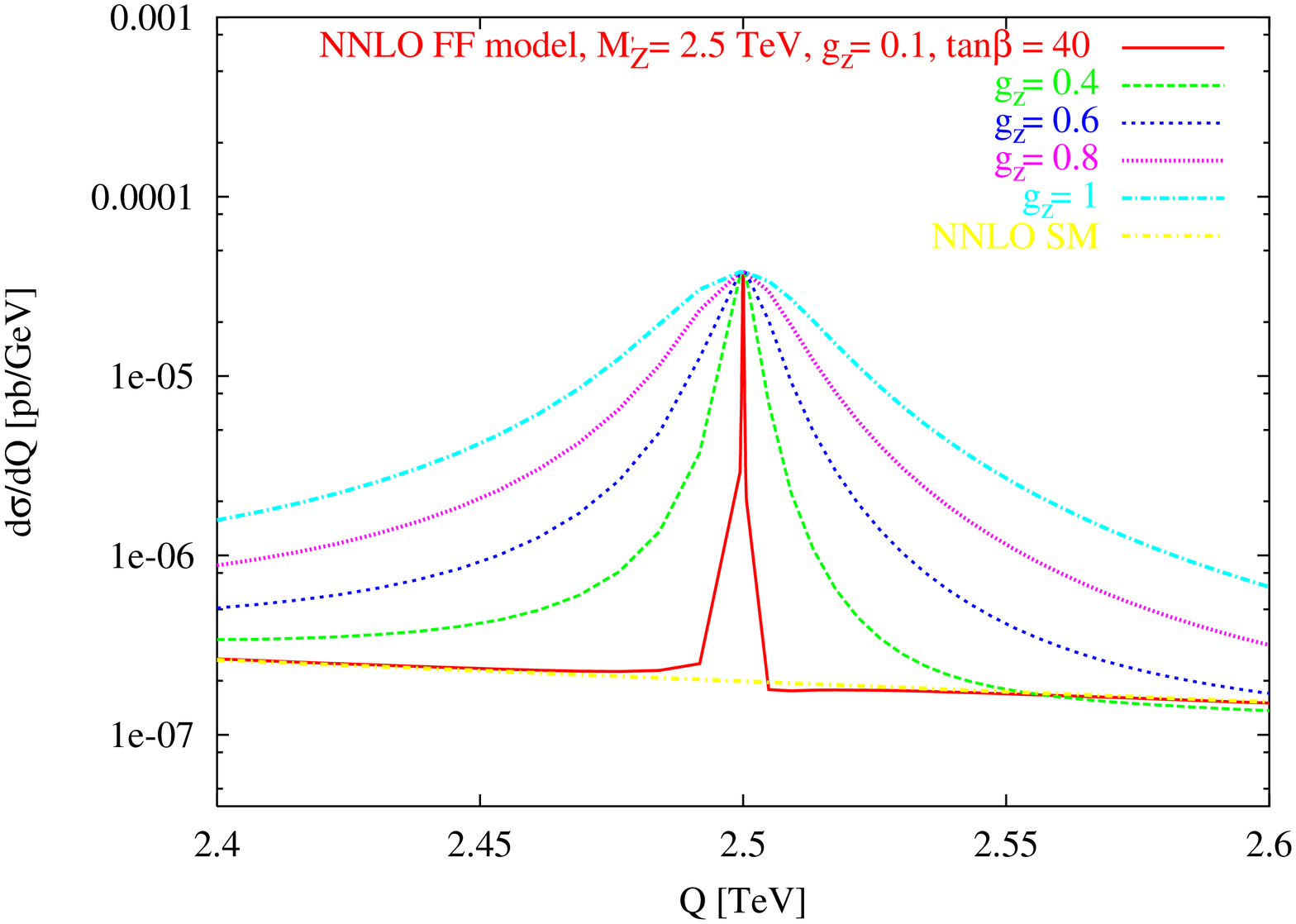}}
\caption{Free fermionic model and the corresponding SM results at NNLO
for $M_{Z^\prime}=2.5$ TeV
for different values of $g_z$ larger than $g_z=0.1$.}
\label{cross012}
\end{figure}

This separation, in general, needs to be done both in
the hard scattering and in the evolution. A zoom of the resonance region
is shown in Fig. \ref{cross09}, which shows that the reduction of the
signal is by a factor of 10 compared to the case
of $M_{Z^\prime}=0.8$ TeV. This drastic reduction of the cross section is
one of the reason why the search of extra neutral currents, if these are
mediated by new gauge bosons of mass above the 1 TeV range, may take
several years of LHC luminosity to be performed, unless the new gauge
coupling is larger. As shown in 
Fig. \ref{cross010}, as we move away from the resonance region, the SM 
background and the  FF result overlap. An interesting feature is that the 
K-factors for the SM result are much larger than for the FF case,
especially as we move from LO to NLO. At NNLO both curves, however, overlap.

\begin{figure}[h]
\subfigure[]{\includegraphics[%
  width=9cm,
  angle=-90]{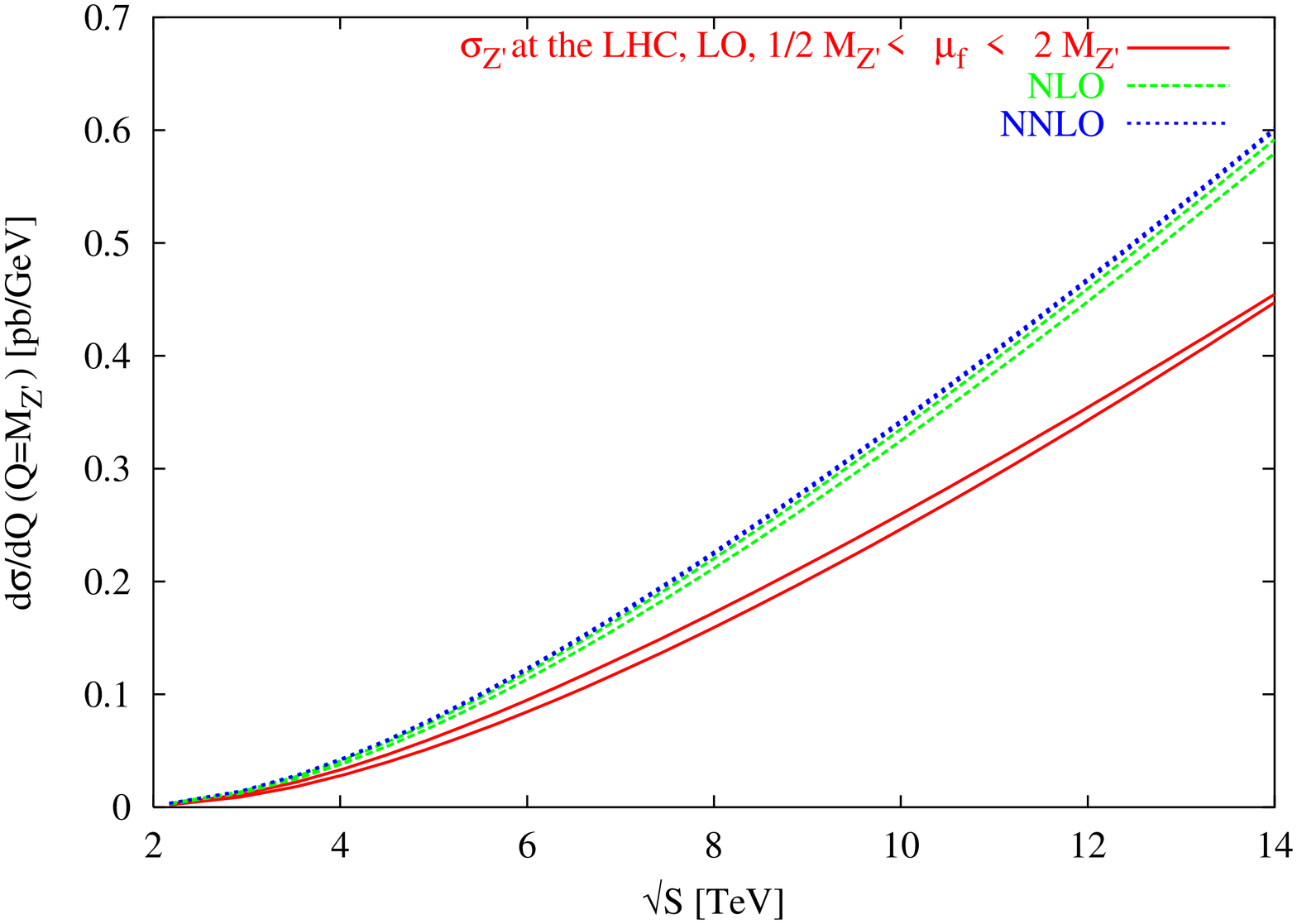}}
\caption{Study of the $\mu_F$ scale dependence in the total cross section
for the $U(1)_{B-L}$ model with $M_{Z^\prime}=0.6$ TeV and $g_z=0.1$.
Here we have chosen $M_{Z^{\prime}}=Q$ for semplicity.}
\label{cross013}
\end{figure}

\begin{figure}[h]
\subfigure[]{\includegraphics[%
  width=9cm,
  angle=-90]{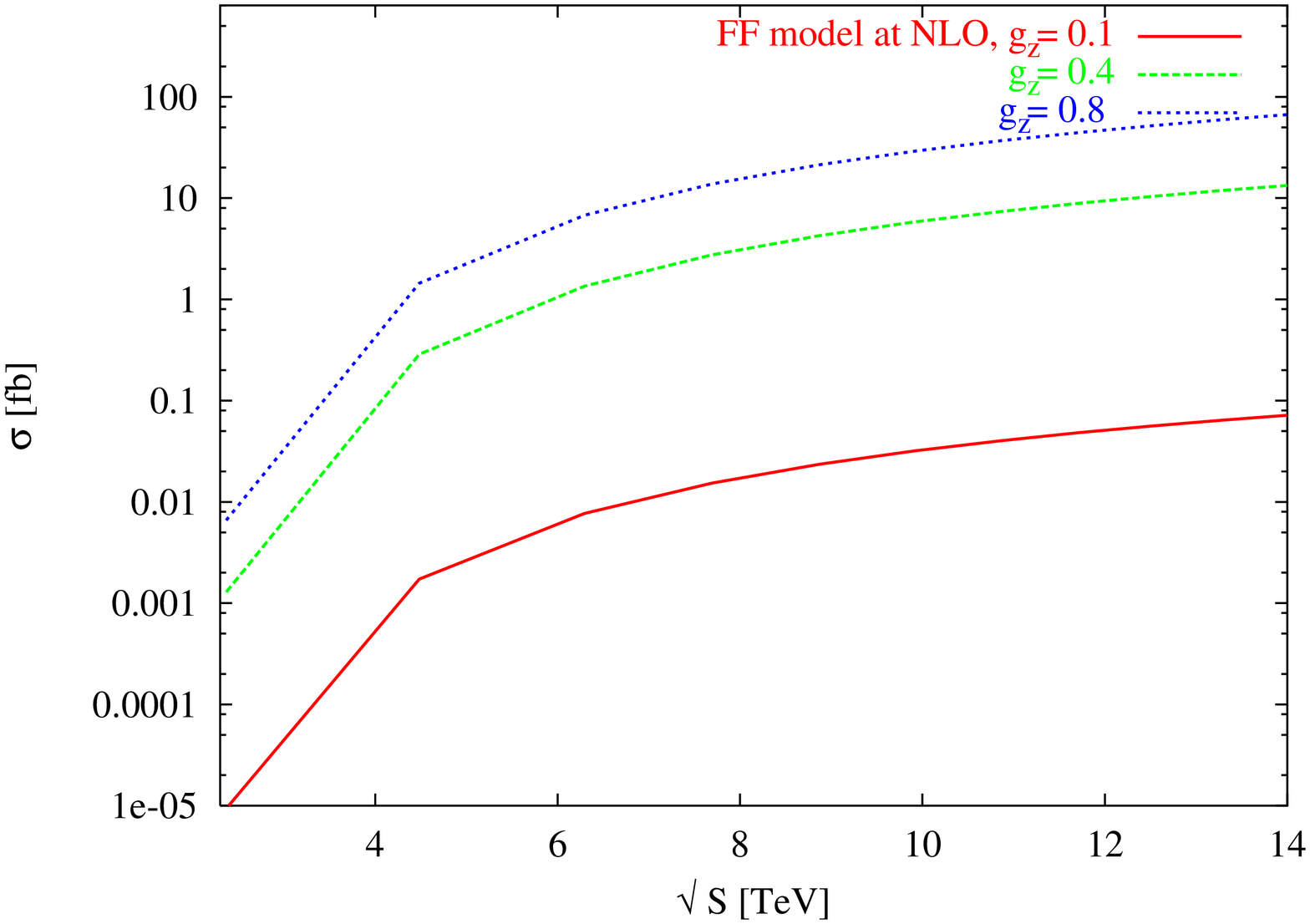}}
\caption{Total cross section for the Free fermionic model at NLO for three different
values of $g_z$ and for $M_{Z^\prime}=1.2$ TeV.
Here we have chosen $\mu_F=\mu_R=Q$ for semplicity and we have integrated the
mass invariant distribution on the interval $M_{Z^{\prime}}\pm 3\Gamma_{Z^{\prime}}$.}
\label{cross014}
\end{figure}

We show, in Fig. \ref{cross011} a plot of the shape of the 
resonance region for $M_{Z^\prime}=2.5$ TeV. The width 
is very narrow (2 GeV) and the size of the cross section down by a factor 
of 100 compared to the case of $M_{Z^\prime}=1.2$ TeV. 
A similar analysis of the shapes of the resonances is shown in
Fig. \ref{cross012} where we have chosen but this time we have varied 
the strength of the new coupling in order to show the widening
of the width, which may easy the detection of the new neutral currents. As shown in
Tab.~(\ref{table_width}), only at large values of the couplings
the size of the width is such to ensure a more direct identification of
the resonance, which should probably be around $30$ GeV or more, in order
not to  be missed.
We conclude this section with the discussion of some results concerning the study
of the variation of the cross section $d\sigma/d Q(Q=M_{Z^\prime})$ (on the peak)
as we vary the factorization scale. In Fig. \ref{cross013} the scale $\mu_f$ has been 
varied in the interval  $1/2 M_{Z^\prime} < \mu_f < 2 M_{Z^\prime}$ for a mass
$M_{Z^\prime}=600$ GeV. These variations are rather small over all the energy interval 
that we have analyzed and show consistently the reduction of the scale dependence of the result moving
from LO to NLO and NNLO. The cross section is sizeable in particular above the 4 TeV scale, 
especially for larger couplings, although the presence of the resonance is not resolved in 
this figure given the
small width. Finally, in Fig. \ref{cross014} we plot the total cross section as a 
function of the energy for 3 values of the new gauge couplings for $M_{Z^\prime}$=1.2 TeV. 
Also in this case the rise of the cross section gets sizeable for larger value of the couplings.

\subsection{NLO/NNLO comparisons and relative differences}
We have included a set of tables which may be useful for actual 
experimental searches and comparisons.
In table \ref{table0m1} we show the LO and in table \ref{table00} the NLO 
results for the invariant mass distributions for the 
first choice (800 GeV) of the mass of the extra 
$Z^\prime$ in all the models, and the corresponding value also for the 
SM. In all the cases the proximity among the
various determinations is quite evident, except on the resonance, where 
the values show a wide variability. The pattern at NNLO, 
shown in table \ref{table01} is similar, and the changes in the cross 
sections from NLO to NNLO in most of the cases are 
around 3 $\%$ or less. These changes are of the same order of those 
obtained by a study of the K-factors in the case of the Z 
resonance \cite{CCG2}. Also for this kinematical region, as on the Z peak 
\cite{CCG2}, the changes from LO to NLO are around 20-30 $\%$, and cover 
the bulk of the QCD corrections. The last several tables describe the 
relative differences between the results of the various models and the 
SM, normalized to the SM values, at the various perturbative orders and 
for 3 values of the coupling constants 
$g_Z=0.05,\,0.1$ and $0.2$.  They give an indication of the role played 
by the changes in the coupling on the behaviour of these observables at 
the tails of the resonance region. In tables \ref{table02} and 
\ref{table03} the region that we explore is between 1 and 1.5 TeV. It is 
rather clear from these results that for a weakly coupled
$Z^\prime$ ($g_Z=0.05$) 
the NLO and NNLO variations respect to the SM result are essentially 
similar. The differences at NLO between the various models and the NLO SM 
are a fraction of a
percent. Therefore, NNLO QCD corrections will not help in this region for 
such weakly coupled extra $Z^\prime$. The differences are not more 
sizeable as we increase the new gauge coupling to 0.1, as shown in 
\ref{table04} and \ref{table05}. Both at NLO and NNLO the difference 
between the SM background and all the other models is smaller than 1 \%. 
Things are not much better for a value of the coupling constant equal to
0.2. The differences between the SM and various models in this region of 
fast fall-off can be of the order of only  2 \%, and just for one model (``$B-L$'').
Given also the small size of these cross sections, which are of the order 
of  $3\times 10^{-2} $ fb, it is hard to separate the various 
contributions. Naturally, the situation will improve considerably if we 
allow a larger gauge coupling since the differences between signal and 
background can become, in principle, quite large.

\begin{table}
\begin{center}
\begin{footnotesize}
\begin{tabular}{|c||c|c|c|}
\hline
\multicolumn{4}{|c|}{$\Gamma_{M_{Z^{\prime}}}(g_z)$ [GeV]}
\tabularnewline
\hline
$g_z$ & $M_{Z^{\prime}}=0.8$ TeV &  $M_{Z^{\prime}}=1.2$ TeV &  $M_{Z^{\prime}}=2.5$ TeV
\tabularnewline
\hline
\hline
 $0.02$   &     $0.004$ & $0.005$ &  $0.012$\tabularnewline\hline
 $0.05$   &     $0.024$ & $0.036$ &  $0.075$\tabularnewline\hline
 $0.1$    &     $0.097$ & $0.146$ &  $0.303$\tabularnewline\hline
 $0.2$    &     $0.388$ & $0.584$ &  $1.215$\tabularnewline\hline
 $0.3$    &     $0.875$ & $1.314$ &  $2.735$\tabularnewline\hline
 $0.4$    &     $1.555$ & $2.336$ &  $4.863$\tabularnewline\hline
 $0.5$    &     $2.430$ & $3.650$ &  $7.598$\tabularnewline\hline
 $0.6$    &     $3.500$ & $5.256$ &  $10.94$\tabularnewline\hline
 $0.7$    &     $4.764$ & $7.154$ &  $14.89$\tabularnewline\hline
 $0.8$    &     $6.223$ & $9.344$ &  $19.45$\tabularnewline\hline
 $0.9$    &     $7.876$ & $11.82$ &  $24.61$\tabularnewline\hline
 $1  $    &     $9.723$ & $14.60$ &  $30.39$\tabularnewline\hline
\end{tabular}
\end{footnotesize}
\end{center}
\caption{Dependence of the total width on the coupling constant $g_z$
for the free fermionic model with $M_{Z^{\prime}}=800$ GeV, $M_{Z^{\prime}}=1.2$ TeV
and $M_{Z^{\prime}}=2.5$ TeV.}
\label{table_width}
\end{table}

\section{Conclusions}
We performed a preliminary comparative analysis of the behaviour of
several models containing extra neutral currents in anomaly-free
constructions and we discussed the implications of the
results for actual experimental searches at the LHC. Compared to other
studies, our objective has been to compare signal and QCD background in a 
series of models, with the highest accuracy, which can be 
systematically performed through NNLO. As expected, the critical 
parameters in order to be able to see a signal of these new interactions 
at the new collider are 
the size of the gauge coupling and the mass of the extra gauge boson, 
while the specific charge assignments of the models play a minor role. 
Other parameters such as $\tan\beta $  also do not play any significant 
role in these types of searches. It is reasonable to believe that much of 
the potentiality for discovering the new resonance, if found, is its 
width, and all the models analyzed so far show very similar 
patterns, with a gauging of
``$B-L$'' being the one that has a slightly wider resonant behaviour. Being
the coupling so important in order to identify which model has better
chances to be confirmed or ruled out, it is necessary, especially
in bottom-up constructions, to rely on more precise investigations
of possible scenarios for the running of the couplings, which are not addressed
in approaches of these types. In the case of
the free fermionic $U(1)$ that we have analyzed, the possibility to 
include these models in a more 
general scenario is natural, since they are naturally produced by a 
unification scheme, but is left for future studies. 
On the other hand, in these and similar models
obtained either in the string picture or in Grand Unification, the 
decoupling of part of the ``extra stuff'' that would complicate the scenario 
that we have analyzed, requires extra assumptions, which would also affect
the running of the couplings of the extra $U(1)'s$. 
These assumptions would introduce various alternatives on the
choice of the symmetry breaking scales, threshold enhancements, 
and so on, which amount, however, to important phenomenological 
details which strongly affect this search. 

Since the V-A structure of the couplings exhibits differences with
respect to other $Z^{\prime}$ models a
measurement of forward-backward asymmetries and/or of charge asymmetries
could be helpful \cite{Petriello:2008zr}, but only if the gauge coupling is sizeable.
The discrimination among the various models remains a very difficult issue for which NNLO QCD determinations, at
least in leptoproduction, though useful, do not seem to be necessary in a first analysis.
For those values of the mass of the extra $Z^\prime$ that we have considered these corrections
cannot be isolated, while the NLO effects remain important.

\clearpage

\centerline{\bf Acknowledgements}
We thank Simone Morelli for discussions and for various cross-checks in
the numerical analysis. M.G. thanks the
Theory Division at the University of Liverpool for hospitality and the
Royal Society for 
financial support. The work of C.C. was supported (in part)
by the European Union through the Marie Curie Research and Training Network
``Universenet'' (MRTN-CT-2006-035863) and by The Interreg II Crete-Cyprus 
Program. He thanks the Theory group at Crete for hospitality.
The work of A.E.F. is supported in part the STFC.

\begin{sidewaystable}
\begin{center}
\begin{footnotesize}
\begin{tabular}{|c||c|c|c|c|c|c|}
\hline
\multicolumn{7}{|c|}{$\textrm{d}\sigma_{LO}/\textrm{d}Q$ [pb/GeV], $M_Z^{\prime}=800$, $g_z=0.1$, $\tan\beta=40$, Candia evol.}
\tabularnewline
\hline
$Q ~[\textrm{GeV}]$&
$\sigma_{LO}(Q)$ FFM &
$\sigma_{LO}(Q)$ \,$U(1)_{B-L}$&
$\sigma_{LO}(Q)$ \,$U(1)_{q+u}$&
$\sigma_{LO}(Q)$ \,$U(1)_{10+\bar{5}}$&
$\sigma_{LO}(Q)$ \,$U(1)_{d-u}$&
$\sigma_{LO}(SM)$\tabularnewline
\hline
\hline
$750$&
$1.1101\cdot10^{-4}$&
$1.0854\cdot10^{-4}$&
$1.0854\cdot10^{-4}$&
$1.1017\cdot10^{-4}$&
$1.1011\cdot10^{-4}$&
$1.1033\cdot10^{-4}$
\tabularnewline
\hline
$761$&
$1.0355\cdot10^{-4}$&
$1.0050\cdot10^{-4}$&
$1.0050\cdot10^{-4}$&
$1.0250\cdot10^{-4}$&
$1.0243\cdot10^{-4}$&
$1.0269\cdot10^{-4}$
\tabularnewline
\hline
$773$&
$9.6852\cdot10^{-5}$&
$9.2759\cdot10^{-5}$&
$9.2759\cdot10^{-5}$&
$9.5421\cdot10^{-5}$&
$9.5315\cdot10^{-5}$&
$9.5674\cdot10^{-5}$
\tabularnewline
\hline
$784$&
$9.1225\cdot10^{-5}$&
$8.4635\cdot10^{-5}$&
$8.4635\cdot10^{-5}$&
$8.8822\cdot10^{-5}$&
$8.8633\cdot10^{-5}$&
$8.9212\cdot10^{-5}$
\tabularnewline
\hline
$796$&
$9.1654\cdot10^{-5}$&
$7.2110\cdot10^{-5}$&
$7.2111\cdot10^{-5}$&
$8.2428\cdot10^{-5}$&
$8.1409\cdot10^{-5}$&
$8.3259\cdot10^{-5}$
\tabularnewline
\hline
$800$&
$1.6448\cdot10^{-2}$&
$4.2388\cdot10^{-2}$&
$2.3928\cdot10^{-2}$&
$1.9570\cdot10^{-2}$&
$4.3085\cdot10^{-2}$&
$8.1086\cdot10^{-5}$
\tabularnewline
\hline
$800$&
$4.9572\cdot10^{-4}$&
$1.9334\cdot10^{-3}$&
$1.8812\cdot10^{-3}$&
$2.8631\cdot10^{-4}$&
$1.5888\cdot10^{-4}$&
$8.0955\cdot10^{-5}$
\tabularnewline
\hline
$801$&
$1.7452\cdot10^{-4}$&
$6.8269\cdot10^{-4}$&
$6.7771\cdot10^{-4}$&
$1.4480\cdot10^{-4}$&
$1.1071\cdot10^{-4}$&
$8.0839\cdot10^{-5}$
\tabularnewline
\hline
$839$&
$6.4010\cdot10^{-5}$&
$6.6355\cdot10^{-5}$&
$6.6355\cdot10^{-5}$&
$6.4761\cdot10^{-5}$&
$6.4800\cdot10^{-5}$&
$6.4607\cdot10^{-5}$
\tabularnewline
\hline
$868$&
$5.4301\cdot10^{-5}$&
$5.5480\cdot10^{-5}$&
$5.5480\cdot10^{-5}$&
$5.4686\cdot10^{-5}$&
$5.4706\cdot10^{-5}$&
$5.4610\cdot10^{-5}$
\tabularnewline
\hline
$900$&
$4.5656\cdot10^{-5}$&
$4.6371\cdot10^{-5}$&
$4.6371\cdot10^{-5}$&
$4.5892\cdot10^{-5}$&
$4.5904\cdot10^{-5}$&
$4.5847\cdot10^{-5}$
\tabularnewline
\hline
\end{tabular}
\end{footnotesize}
\caption{LO invariant mass distributions}
\label{table0m1}
\end{center}
\end{sidewaystable}

\begin{sidewaystable}
\begin{center}
\begin{footnotesize}
\begin{tabular}{|c||c|c|c|c|c|c|}
\hline
\multicolumn{7}{|c|}{$\textrm{d}\sigma_{NLO}/\textrm{d}Q$ [pb/GeV], $M_Z^{\prime}=800$, $g_z=0.1$, $\tan\beta=40$, Candia evol.}
\tabularnewline
\hline
$Q ~[\textrm{GeV}]$&
$\sigma_{NLO}(Q)$ FFM &
$\sigma_{NLO}(Q)$ \,$U(1)_{B-L}$&
$\sigma_{NLO}(Q)$ \,$U(1)_{q+u}$&
$\sigma_{NLO}(Q)$ \,$U(1)_{10+\bar{5}}$&
$\sigma_{NLO}(Q)$ \,$U(1)_{d-u}$&
$\sigma_{NLO}(SM)$\tabularnewline
\hline
\hline
$750$&
$1.4362\cdot10^{-4}$&
$1.4048\cdot10^{-4}$&
$1.4048\cdot10^{-4}$&
$1.4257\cdot10^{-4}$&
$1.4248\cdot10^{-4}$&
$1.4276\cdot10^{-4}$
\tabularnewline
\hline
$761$&
$1.3394\cdot10^{-4}$&
$1.3008\cdot10^{-4}$&
$1.3008\cdot10^{-4}$&
$1.3263\cdot10^{-4}$&
$1.3252\cdot10^{-4}$&
$1.3287\cdot10^{-4}$
\tabularnewline
\hline
$773$&
$1.2526\cdot10^{-4}$&
$1.2006\cdot10^{-4}$&
$1.2006\cdot10^{-4}$&
$1.2346\cdot10^{-4}$&
$1.2331\cdot10^{-4}$&
$1.2377\cdot10^{-4}$
\tabularnewline
\hline
$784$&
$1.1794\cdot10^{-4}$&
$1.0958\cdot10^{-4}$&
$1.0958\cdot10^{-4}$&
$1.1492\cdot10^{-4}$&
$1.1465\cdot10^{-4}$&
$1.1540\cdot10^{-4}$
\tabularnewline
\hline
$796$&
$1.1834\cdot10^{-4}$&
$9.3647\cdot10^{-5}$&
$9.3648\cdot10^{-5}$&
$1.0671\cdot10^{-4}$&
$1.0530\cdot10^{-4}$&
$1.0769\cdot10^{-4}$
\tabularnewline
\hline
$800$&
$2.1411\cdot10^{-2}$&
$5.4992\cdot10^{-2}$&
$3.1043\cdot10^{-2}$&
$2.5389\cdot10^{-2}$&
$5.5896\cdot10^{-2}$&
$1.0487\cdot10^{-4}$
\tabularnewline
\hline
$800$&
$6.4893\cdot10^{-4}$&
$2.5014\cdot10^{-3}$&
$2.4339\cdot10^{-3}$&
$3.6947\cdot10^{-4}$&
$2.0566\cdot10^{-4}$&
$1.0470\cdot10^{-4}$
\tabularnewline
\hline
$801$&
$2.2882\cdot10^{-4}$&
$8.8180\cdot10^{-4}$&
$8.7537\cdot10^{-4}$&
$1.8666\cdot10^{-4}$&
$1.4323\cdot10^{-4}$&
$1.0455\cdot10^{-4}$
\tabularnewline
\hline
$839$&
$8.2772\cdot10^{-5}$&
$8.5749\cdot10^{-5}$&
$8.5749\cdot10^{-5}$&
$8.3712\cdot10^{-5}$&
$8.3771\cdot10^{-5}$&
$8.3523\cdot10^{-5}$
\tabularnewline
\hline
$868$&
$7.0183\cdot10^{-5}$&
$7.1679\cdot10^{-5}$&
$7.1679\cdot10^{-5}$&
$7.0666\cdot10^{-5}$&
$7.0696\cdot10^{-5}$&
$7.0573\cdot10^{-5}$
\tabularnewline
\hline
$900$&
$5.8982\cdot10^{-5}$&
$5.9888\cdot10^{-5}$&
$5.9888\cdot10^{-5}$&
$5.9278\cdot10^{-5}$&
$5.9296\cdot10^{-5}$&
$5.9222\cdot10^{-5}$
\tabularnewline
\hline
\end{tabular}
\end{footnotesize}
\end{center}
\caption{NLO distributions for $750 < Q < 900$ GeV }
\label{table00}

\begin{center}
\begin{footnotesize}
\begin{tabular}{|c||c|c|c|c|c|c|}
\hline
\multicolumn{7}{|c|}{$\textrm{d}\sigma_{NNLO}/\textrm{d}Q$ [pb/GeV], $M_Z^{\prime}=800$, $g_z=0.1$, $\tan\beta=40$, Candia evol.}
\tabularnewline
\hline
$Q ~[\textrm{GeV}]$&
$\sigma_{NNLO}(Q)$ FFM &
$\sigma_{NNLO}(Q)$ \,$U(1)_{B-L}$&
$\sigma_{NNLO}(Q)$ \,$U(1)_{q+u}$&
$\sigma_{NNLO}(Q)$ \,$U(1)_{10+\bar{5}}$&
$\sigma_{NNLO}(Q)$ \,$U(1)_{d-u}$&
$\sigma_{NNLO}(SM)$\tabularnewline
\hline
\hline
$750$&
$1.4793\cdot10^{-4}$&
$1.4472\cdot10^{-4}$&
$1.4472\cdot10^{-4}$&
$1.4686\cdot10^{-4}$&
$1.4676\cdot10^{-4}$&
$1.4705\cdot10^{-4}$
\tabularnewline
\hline
$761$&
$1.3803\cdot10^{-4}$&
$1.3407\cdot10^{-4}$&
$1.3407\cdot10^{-4}$&
$1.3669\cdot10^{-4}$&
$1.3657\cdot10^{-4}$&
$1.3693\cdot10^{-4}$
\tabularnewline
\hline
$773$&
$1.2914\cdot10^{-4}$&
$1.2382\cdot10^{-4}$&
$1.2382\cdot10^{-4}$&
$1.2730\cdot10^{-4}$&
$1.2714\cdot10^{-4}$&
$1.2762\cdot10^{-4}$
\tabularnewline
\hline
$784$&
$1.2164\cdot10^{-4}$&
$1.1308\cdot10^{-4}$&
$1.1308\cdot10^{-4}$&
$1.1856\cdot10^{-4}$&
$1.1827\cdot10^{-4}$&
$1.1904\cdot10^{-4}$
\tabularnewline
\hline
$796$&
$1.2207\cdot10^{-4}$&
$9.6772\cdot10^{-5}$&
$9.6773\cdot10^{-5}$&
$1.1017\cdot10^{-4}$&
$1.0867\cdot10^{-4}$&
$1.1114\cdot10^{-4}$
\tabularnewline
\hline
$800$&
$2.2140\cdot10^{-2}$&
$5.6805\cdot10^{-2}$&
$3.2066\cdot10^{-2}$&
$2.6233\cdot10^{-2}$&
$5.7755\cdot10^{-2}$&
$1.0825\cdot10^{-4}$
\tabularnewline
\hline
$800$&
$6.7222\cdot10^{-4}$&
$2.5818\cdot10^{-3}$&
$2.5121\cdot10^{-3}$&
$3.8114\cdot10^{-4}$&
$2.1235\cdot10^{-4}$&
$1.0808\cdot10^{-4}$
\tabularnewline
\hline
$801$&
$2.3717\cdot10^{-4}$&
$9.0971\cdot10^{-4}$&
$9.0307\cdot10^{-4}$&
$1.9249\cdot10^{-4}$&
$1.4787\cdot10^{-4}$&
$1.0793\cdot10^{-4}$
\tabularnewline
\hline
$839$&
$8.5581\cdot10^{-5}$&
$8.8638\cdot10^{-5}$&
$8.8638\cdot10^{-5}$&
$8.6542\cdot10^{-5}$&
$8.6606\cdot10^{-5}$&
$8.6349\cdot10^{-5}$
\tabularnewline
\hline
$868$&
$7.2645\cdot10^{-5}$&
$7.4182\cdot10^{-5}$&
$7.4182\cdot10^{-5}$&
$7.3139\cdot10^{-5}$&
$7.3171\cdot10^{-5}$&
$7.3044\cdot10^{-5}$
\tabularnewline
\hline
$900$&
$6.1122\cdot10^{-5}$&
$6.2054\cdot10^{-5}$&
$6.2054\cdot10^{-5}$&
$6.1425\cdot10^{-5}$&
$6.1444\cdot10^{-5}$&
$6.1368\cdot10^{-5}$
\tabularnewline
\hline
\end{tabular}
\end{footnotesize}
\end{center}
\caption{NNLO distributions for $750 < Q < 900$ GeV}
\label{table01}
\end{sidewaystable}

\begin{sidewaystable}
\begin{center}
\begin{footnotesize}
\begin{tabular}{|c||c|c|c|c|c|c|}
\hline
\multicolumn{7}{|c|}{$|\sigma_{nlo}^{SM}-\sigma_{nlo}^{i}|/\sigma_{nlo}^{SM}\%$ , $M_Z^{\prime}=800$, $g_z=0.05$, $\tan\beta=40$, Candia evol.}
\tabularnewline
\hline
$Q ~[\textrm{GeV}]$&
$\sigma_{nlo}^{SM}(Q)$[pb/GeV]&
$\Delta_{nlo}^{FFM}\%$  &
$\Delta_{nlo}^{B-L}\%$  &
$\Delta_{nlo}^{q+u}\%$  &
$\Delta_{nlo}^{10+\bar{5}}\%$&
$\Delta_{nlo}^{d-u}\%$\tabularnewline
\hline
\hline
$1000$&
$3.5146\cdot10^{-5}$&
$6.5325\cdot10^{-2}$&
$1.6003\cdot10^{-1}$&
$1.6003\cdot10^{-1}$&
$1.0162\cdot10^{-2}$&
$1.4126\cdot10^{-2}$
\tabularnewline
\hline
$1015$&
$3.2618\cdot10^{-5}$&
$6.2528\cdot10^{-2}$&
$1.5220\cdot10^{-1}$&
$1.5220\cdot10^{-1}$&
$9.5155\cdot10^{-3}$&
$1.3203\cdot10^{-2}$
\tabularnewline
\hline
$1030$&
$3.0299\cdot10^{-5}$&
$6.0105\cdot10^{-2}$&
$1.4541\cdot10^{-1}$&
$1.4541\cdot10^{-1}$&
$8.9574\cdot10^{-3}$&
$1.2402\cdot10^{-2}$
\tabularnewline
\hline
$1045$&
$2.8168\cdot10^{-5}$&
$5.7987\cdot10^{-2}$&
$1.3947\cdot10^{-1}$&
$1.3947\cdot10^{-1}$&
$8.4716\cdot10^{-3}$&
$1.1701\cdot10^{-2}$
\tabularnewline
\hline
$1060$&
$2.6209\cdot10^{-5}$&
$5.6121\cdot10^{-2}$&
$1.3424\cdot10^{-1}$&
$1.3424\cdot10^{-1}$&
$8.0455\cdot10^{-3}$&
$1.1083\cdot10^{-2}$
\tabularnewline
\hline
$1165$&
$1.6156\cdot10^{-5}$&
$4.7509\cdot10^{-2}$&
$1.1003\cdot10^{-1}$&
$1.1003\cdot10^{-1}$&
$6.1091\cdot10^{-3}$&
$8.2061\cdot10^{-3}$
\tabularnewline
\hline
$1210$&
$1.3265\cdot10^{-5}$&
$4.5240\cdot10^{-2}$&
$1.0361\cdot10^{-1}$&
$1.0361\cdot10^{-1}$&
$5.6126\cdot10^{-3}$&
$7.4369\cdot10^{-3}$
\tabularnewline
\hline
$1250$&
$1.1183\cdot10^{-5}$&
$4.3636\cdot10^{-2}$&
$9.9050\cdot10^{-2}$&
$9.9050\cdot10^{-2}$&
$5.2687\cdot10^{-3}$&
$6.8881\cdot10^{-3}$
\tabularnewline
\hline
$1355$&
$7.2763\cdot10^{-6}$&
$4.0639\cdot10^{-2}$&
$9.0453\cdot10^{-2}$&
$9.0453\cdot10^{-2}$&
$4.6479\cdot10^{-3}$&
$5.8404\cdot10^{-3}$
\tabularnewline
\hline
$1425$&
$5.5361\cdot10^{-6}$&
$3.9279\cdot10^{-2}$&
$8.6492\cdot10^{-2}$&
$8.6492\cdot10^{-2}$&
$4.3830\cdot10^{-3}$&
$5.3492\cdot10^{-3}$
\tabularnewline
\hline
$1500$&
$4.1734\cdot10^{-6}$&
$3.8186\cdot10^{-2}$&
$8.3253\cdot10^{-2}$&
$8.3253\cdot10^{-2}$&
$4.1841\cdot10^{-3}$&
$4.9401\cdot10^{-3}$
\tabularnewline
\hline
\end{tabular}
\end{footnotesize}
\caption{Percentage differences at NLO. We define $\Delta_{nlo}^{i}=|\sigma_{nlo}^{SM}-\sigma_{nlo}^{i}|/\sigma_{nlo}^{SM}$ where $i=FFM,B-L,q+u,10+\bar{5},d-u$.}
\label{table02}
\end{center}
\begin{center}
\begin{footnotesize}
\begin{tabular}{|c||c|c|c|c|c|c|}
\hline
\multicolumn{7}{|c|}{$|\sigma_{nnlo}^{SM}-\sigma_{nnlo}^{i}|/\sigma_{nnlo}^{SM}\%$ , $M_Z^{\prime}=800$, $g_z=0.05$, $\tan\beta=40$, Candia evol.}
\tabularnewline
\hline
$Q ~[\textrm{GeV}]$&
$\sigma_{nnlo}^{SM}(Q)$[pb/GeV]&
$\Delta_{nnlo}^{FFM}\%$  &
$\Delta_{nnlo}^{B-L}\%$  &
$\Delta_{nnlo}^{q+u}\%$  &
$\Delta_{nnlo}^{10+\bar{5}}\%$&
$\Delta_{nnlo}^{d-u}\%$\tabularnewline
\hline
\hline
$1000$&
$3.6546\cdot10^{-5}$&
$6.4565\cdot10^{-2}$&
$1.5879\cdot10^{-1}$&
$1.5879\cdot10^{-1}$&
$9.8298\cdot10^{-3}$&
$1.4114\cdot10^{-2}$
\tabularnewline
\hline
$1015$&
$3.3935\cdot10^{-5}$&
$6.1789\cdot10^{-2}$&
$1.5099\cdot10^{-1}$&
$1.5099\cdot10^{-1}$&
$9.1914\cdot10^{-3}$&
$1.3191\cdot10^{-2}$
\tabularnewline
\hline
$1030$&
$3.1537\cdot10^{-5}$&
$5.9383\cdot10^{-2}$&
$1.4423\cdot10^{-1}$&
$1.4423\cdot10^{-1}$&
$8.6399\cdot10^{-3}$&
$1.2390\cdot10^{-2}$
\tabularnewline
\hline
$1045$&
$2.9334\cdot10^{-5}$&
$5.7279\cdot10^{-2}$&
$1.3831\cdot10^{-1}$&
$1.3831\cdot10^{-1}$&
$8.1595\cdot10^{-3}$&
$1.1689\cdot10^{-2}$
\tabularnewline
\hline
$1060$&
$2.7306\cdot10^{-5}$&
$5.5425\cdot10^{-2}$&
$1.3310\cdot10^{-1}$&
$1.3310\cdot10^{-1}$&
$7.7381\cdot10^{-3}$&
$1.1071\cdot10^{-2}$
\tabularnewline
\hline
$1165$&
$1.6888\cdot10^{-5}$&
$4.6857\cdot10^{-2}$&
$1.0895\cdot10^{-1}$&
$1.0895\cdot10^{-1}$&
$5.8167\cdot10^{-3}$&
$8.1924\cdot10^{-3}$
\tabularnewline
\hline
$1210$&
$1.3884\cdot10^{-5}$&
$4.4594\cdot10^{-2}$&
$1.0254\cdot10^{-1}$&
$1.0254\cdot10^{-1}$&
$5.3213\cdot10^{-3}$&
$7.4223\cdot10^{-3}$
\tabularnewline
\hline
$1250$&
$1.1718\cdot10^{-5}$&
$4.2991\cdot10^{-2}$&
$9.7986\cdot10^{-2}$&
$9.7986\cdot10^{-2}$&
$4.9768\cdot10^{-3}$&
$6.8729\cdot10^{-3}$
\tabularnewline
\hline
$1355$&
$7.6472\cdot10^{-6}$&
$3.9988\cdot10^{-2}$&
$8.9375\cdot10^{-2}$&
$8.9375\cdot10^{-2}$&
$4.3502\cdot10^{-3}$&
$5.8236\cdot10^{-3}$
\tabularnewline
\hline
$1425$&
$5.8293\cdot10^{-6}$&
$3.8618\cdot10^{-2}$&
$8.5395\cdot10^{-2}$&
$8.5395\cdot10^{-2}$&
$4.0792\cdot10^{-3}$&
$5.3312\cdot10^{-3}$
\tabularnewline
\hline
$1500$&
$4.4031\cdot10^{-6}$&
$3.7510\cdot10^{-2}$&
$8.2129\cdot10^{-2}$&
$8.2129\cdot10^{-2}$&
$3.8720\cdot10^{-3}$&
$4.9211\cdot10^{-3}$
\tabularnewline
\hline
\end{tabular}
\end{footnotesize}
\end{center}
\caption{Percentage differences at NNLO. We define $\Delta_{nnlo}^{i}=|\sigma_{nnlo}^{SM}-\sigma_{nnlo}^{i}|/\sigma_{nnlo}^{SM}$.}
\label{table03}
\end{sidewaystable}

\begin{sidewaystable}
\begin{center}
\begin{footnotesize}
\begin{tabular}{|c||c|c|c|c|c|c|}
\hline
\multicolumn{7}{|c|}{$|\sigma_{nlo}^{SM}-\sigma_{nlo}^{i}|/\sigma_{nlo}^{SM}\%$ , $M_Z^{\prime}=800$, $g_z=0.1$, $\tan\beta=40$, Candia evol.}
\tabularnewline
\hline
$Q ~[\textrm{GeV}]$&
$\sigma_{nlo}^{SM}(Q)$[pb/GeV]&
$\Delta_{nlo}^{FFM}\%$  &
$\Delta_{nlo}^{B-L}\%$  &
$\Delta_{nlo}^{q+u}\%$  &
$\Delta_{nlo}^{10+\bar{5}}\%$&
$\Delta_{nlo}^{d-u}\%$\tabularnewline
\hline
\hline
$1000$&
$3.5146\cdot10^{-5}$&
$2.4555\cdot10^{-1}$&
$6.5950\cdot10^{-1}$&
$6.5950\cdot10^{-1}$&
$5.5268\cdot10^{-2}$&
$7.0677\cdot10^{-2}$
\tabularnewline
\hline
$1015$&
$3.2618\cdot10^{-5}$&
$2.3454\cdot10^{-1}$&
$6.2764\cdot10^{-1}$&
$6.2764\cdot10^{-1}$&
$5.2612\cdot10^{-2}$&
$6.6957\cdot10^{-2}$
\tabularnewline
\hline
$1030$&
$3.0299\cdot10^{-5}$&
$2.2500\cdot10^{-1}$&
$6.0005\cdot10^{-1}$&
$6.0005\cdot10^{-1}$&
$5.0320\cdot10^{-2}$&
$6.3730\cdot10^{-2}$
\tabularnewline
\hline
$1045$&
$2.8168\cdot10^{-5}$&
$2.1666\cdot10^{-1}$&
$5.7593\cdot10^{-1}$&
$5.7593\cdot10^{-1}$&
$4.8324\cdot10^{-2}$&
$6.0904\cdot10^{-2}$
\tabularnewline
\hline
$1060$&
$2.6209\cdot10^{-5}$&
$2.0931\cdot10^{-1}$&
$5.5469\cdot10^{-1}$&
$5.5469\cdot10^{-1}$&
$4.6574\cdot10^{-2}$&
$5.8410\cdot10^{-2}$
\tabularnewline
\hline
$1165$&
$1.6156\cdot10^{-5}$&
$1.7534\cdot10^{-1}$&
$4.5649\cdot10^{-1}$&
$4.5649\cdot10^{-1}$&
$3.8610\cdot10^{-2}$&
$4.6791\cdot10^{-2}$
\tabularnewline
\hline
$1210$&
$1.3265\cdot10^{-5}$&
$1.6639\cdot10^{-1}$&
$4.3048\cdot10^{-1}$&
$4.3048\cdot10^{-1}$&
$3.6562\cdot10^{-2}$&
$4.3677\cdot10^{-2}$
\tabularnewline
\hline
$1250$&
$1.1183\cdot10^{-5}$&
$1.6007\cdot10^{-1}$&
$4.1202\cdot10^{-1}$&
$4.1202\cdot10^{-1}$&
$3.5137\cdot10^{-2}$&
$4.1449\cdot10^{-2}$
\tabularnewline
\hline
$1355$&
$7.2763\cdot10^{-6}$&
$1.4826\cdot10^{-1}$&
$3.7722\cdot10^{-1}$&
$3.7722\cdot10^{-1}$&
$3.2554\cdot10^{-2}$&
$3.7189\cdot10^{-2}$
\tabularnewline
\hline
$1425$&
$5.5361\cdot10^{-6}$&
$1.4291\cdot10^{-1}$&
$3.6118\cdot10^{-1}$&
$3.6118\cdot10^{-1}$&
$3.1440\cdot10^{-2}$&
$3.5182\cdot10^{-2}$
\tabularnewline
\hline
$1500$&
$4.1734\cdot10^{-6}$&
$1.3861\cdot10^{-1}$&
$3.4806\cdot10^{-1}$&
$3.4806\cdot10^{-1}$&
$3.0591\cdot10^{-2}$&
$3.3504\cdot10^{-2}$
\tabularnewline
\hline
\end{tabular}
\end{footnotesize}
\caption{Percentage differences at NLO for $g_Z=0.1$. Here and in the following we 
use the same notation of the previous tables.}
\label{table04}
\end{center}
\hspace{-1cm}
\begin{center}
\begin{footnotesize}
\begin{tabular}{|c||c|c|c|c|c|c|}
\hline
\multicolumn{7}{|c|}{$|\sigma_{nnlo}^{SM}-\sigma_{nnlo}^{i}|/\sigma_{nnlo}^{SM}\%$ , $M_Z^{\prime}=800$, $g_z=0.1$, $\tan\beta=40$, Candia evol.}
\tabularnewline
\hline
$Q ~[\textrm{GeV}]$&
$\sigma_{nnlo}^{SM}(Q)$[pb/GeV]&
$\Delta_{nnlo}^{FFM}\%$  &
$\Delta_{nnlo}^{B-L}\%$  &
$\Delta_{nnlo}^{q+u}\%$  &
$\Delta_{nnlo}^{10+\bar{5}}\%$&
$\Delta_{nnlo}^{d-u}\%$\tabularnewline
\hline
\hline
$1000$&
$3.6546\cdot10^{-5}$&
$2.4248\cdot10^{-1}$&
$6.5458\cdot10^{-1}$&
$6.5458\cdot10^{-1}$&
$5.3970\cdot10^{-2}$&
$7.0662\cdot10^{-2}$
\tabularnewline
\hline
$1015$&
$3.3935\cdot10^{-5}$&
$2.3155\cdot10^{-1}$&
$6.2284\cdot10^{-1}$&
$6.2284\cdot10^{-1}$&
$5.1346\cdot10^{-2}$&
$6.6942\cdot10^{-2}$
\tabularnewline
\hline
$1030$&
$3.1537\cdot10^{-5}$&
$2.2208\cdot10^{-1}$&
$5.9536\cdot10^{-1}$&
$5.9536\cdot10^{-1}$&
$4.9081\cdot10^{-2}$&
$6.3715\cdot10^{-2}$
\tabularnewline
\hline
$1045$&
$2.9334\cdot10^{-5}$&
$2.1379\cdot10^{-1}$&
$5.7133\cdot10^{-1}$&
$5.7133\cdot10^{-1}$&
$4.7108\cdot10^{-2}$&
$6.0888\cdot10^{-2}$
\tabularnewline
\hline
$1060$&
$2.7306\cdot10^{-5}$&
$2.0649\cdot10^{-1}$&
$5.5017\cdot10^{-1}$&
$5.5017\cdot10^{-1}$&
$4.5376\cdot10^{-2}$&
$5.8394\cdot10^{-2}$
\tabularnewline
\hline
$1165$&
$1.6888\cdot10^{-5}$&
$1.7269\cdot10^{-1}$&
$4.5223\cdot10^{-1}$&
$4.5223\cdot10^{-1}$&
$3.7478\cdot10^{-2}$&
$4.6774\cdot10^{-2}$
\tabularnewline
\hline
$1210$&
$1.3884\cdot10^{-5}$&
$1.6377\cdot10^{-1}$&
$4.2627\cdot10^{-1}$&
$4.2627\cdot10^{-1}$&
$3.5438\cdot10^{-2}$&
$4.3659\cdot10^{-2}$
\tabularnewline
\hline
$1250$&
$1.1718\cdot10^{-5}$&
$1.5744\cdot10^{-1}$&
$4.0781\cdot10^{-1}$&
$4.0781\cdot10^{-1}$&
$3.4013\cdot10^{-2}$&
$4.1431\cdot10^{-2}$
\tabularnewline
\hline
$1355$&
$7.6472\cdot10^{-6}$&
$1.4560\cdot10^{-1}$&
$3.7296\cdot10^{-1}$&
$3.7296\cdot10^{-1}$&
$3.1411\cdot10^{-2}$&
$3.7169\cdot10^{-2}$
\tabularnewline
\hline
$1425$&
$5.8293\cdot10^{-6}$&
$1.4021\cdot10^{-1}$&
$3.5685\cdot10^{-1}$&
$3.5685\cdot10^{-1}$&
$3.0275\cdot10^{-2}$&
$3.5160\cdot10^{-2}$
\tabularnewline
\hline
$1500$&
$4.4031\cdot10^{-6}$&
$1.3585\cdot10^{-1}$&
$3.4362\cdot10^{-1}$&
$3.4362\cdot10^{-1}$&
$2.9397\cdot10^{-2}$&
$3.3481\cdot10^{-2}$
\tabularnewline
\hline
\end{tabular}
\end{footnotesize}
\end{center}
\caption{Percentage differences at NNLO for $g_Z=0.1$}
\label{table05}
\end{sidewaystable}

\begin{sidewaystable}
\begin{center}
\begin{footnotesize}
\begin{tabular}{|c||c|c|c|c|c|c|}
\hline
\multicolumn{7}{|c|}{$|\sigma_{nlo}^{SM}-\sigma_{nlo}^{i}|/\sigma_{nlo}^{SM}\%$ , $M_Z^{\prime}=800$, $g_z=0.2$, $\tan\beta=40$, Candia evol.}
\tabularnewline
\hline
$Q ~[\textrm{GeV}]$&
$\sigma_{nlo}^{SM}(Q)$[pb/GeV]&
$\Delta_{nlo}^{FFM}\%$  &
$\Delta_{nlo}^{B-L}\%$  &
$\Delta_{nlo}^{q+u}\%$  &
$\Delta_{nlo}^{10+\bar{5}}\%$&
$\Delta_{nlo}^{d-u}\%$\tabularnewline
\hline
\hline
$1000$&
$3.5146\cdot10^{-5}$&
$9.4061\cdot10^{-1}$&
$2.7377\cdot10^{+0}$&
$2.7377\cdot10^{+0}$&
$2.4462\cdot10^{-1}$&
$2.9911\cdot10^{-1}$
\tabularnewline
\hline
$1015$&
$3.2618\cdot10^{-5}$&
$8.9927\cdot10^{-1}$&
$2.6020\cdot10^{+0}$&
$2.6020\cdot10^{+0}$&
$2.3306\cdot10^{-1}$&
$2.8399\cdot10^{-1}$
\tabularnewline
\hline
$1030$&
$3.0299\cdot10^{-5}$&
$8.6334\cdot10^{-1}$&
$2.4847\cdot10^{+0}$&
$2.4847\cdot10^{+0}$&
$2.2312\cdot10^{-1}$&
$2.7088\cdot10^{-1}$
\tabularnewline
\hline
$1045$&
$2.8168\cdot10^{-5}$&
$8.3183\cdot10^{-1}$&
$2.3825\cdot10^{+0}$&
$2.3825\cdot10^{+0}$&
$2.1448\cdot10^{-1}$&
$2.5940\cdot10^{-1}$
\tabularnewline
\hline
$1060$&
$2.6209\cdot10^{-5}$&
$8.0401\cdot10^{-1}$&
$2.2926\cdot10^{+0}$&
$2.2926\cdot10^{+0}$&
$2.0693\cdot10^{-1}$&
$2.4928\cdot10^{-1}$
\tabularnewline
\hline
$1165$&
$1.6156\cdot10^{-5}$&
$6.7477\cdot10^{-1}$&
$1.8795\cdot10^{+0}$&
$1.8795\cdot10^{+0}$&
$1.7274\cdot10^{-1}$&
$2.0217\cdot10^{-1}$
\tabularnewline
\hline
$1210$&
$1.3265\cdot10^{-5}$&
$6.4052\cdot10^{-1}$&
$1.7707\cdot10^{+0}$&
$1.7707\cdot10^{+0}$&
$1.6400\cdot10^{-1}$&
$1.8955\cdot10^{-1}$
\tabularnewline
\hline
$1250$&
$1.1183\cdot10^{-5}$&
$6.1627\cdot10^{-1}$&
$1.6937\cdot10^{+0}$&
$1.6937\cdot10^{+0}$&
$1.5792\cdot10^{-1}$&
$1.8052\cdot10^{-1}$
\tabularnewline
\hline
$1355$&
$7.2763\cdot10^{-6}$&
$5.7094\cdot10^{-1}$&
$1.5487\cdot10^{+0}$&
$1.5487\cdot10^{+0}$&
$1.4689\cdot10^{-1}$&
$1.6326\cdot10^{-1}$
\tabularnewline
\hline
$1425$&
$5.5361\cdot10^{-6}$&
$5.5039\cdot10^{-1}$&
$1.4820\cdot10^{+0}$&
$1.4820\cdot10^{+0}$&
$1.4212\cdot10^{-1}$&
$1.5512\cdot10^{-1}$
\tabularnewline
\hline
$1500$&
$4.1734\cdot10^{-6}$&
$5.3389\cdot10^{-1}$&
$1.4275\cdot10^{+0}$&
$1.4275\cdot10^{+0}$&
$1.3846\cdot10^{-1}$&
$1.4832\cdot10^{-1}$
\tabularnewline
\hline
\end{tabular}
\end{footnotesize}
\caption{Percentage differences at NLO for $g_Z=0.2$}
\label{table06}
\end{center}
\hspace{-1cm}
\begin{center}
\begin{footnotesize}
\begin{tabular}{|c||c|c|c|c|c|c|}
\hline
\multicolumn{7}{|c|}{$|\sigma_{nnlo}^{SM}-\sigma_{nnlo}^{i}|/\sigma_{nnlo}^{SM}\%$ , $M_Z^{\prime}=800$, $g_z=0.2$, $\tan\beta=40$, Candia evol.}
\tabularnewline
\hline
$Q ~[\textrm{GeV}]$&
$\sigma_{nnlo}^{SM}(Q)$[pb/GeV]&
$\Delta_{nnlo}^{FFM}\%$  &
$\Delta_{nnlo}^{B-L}\%$  &
$\Delta_{nnlo}^{q+u}\%$  &
$\Delta_{nnlo}^{10+\bar{5}}\%$&
$\Delta_{nnlo}^{d-u}\%$\tabularnewline
\hline
\hline
$1000$&
$3.6546\cdot10^{-5}$&
$9.2821\cdot10^{-1}$&
$2.7182\cdot10^{+0}$&
$2.7182\cdot10^{+0}$&
$2.3947\cdot10^{-1}$&
$2.9909\cdot10^{-1}$
\tabularnewline
\hline
$1015$&
$3.3935\cdot10^{-5}$&
$8.8720\cdot10^{-1}$&
$2.5829\cdot10^{+0}$&
$2.5829\cdot10^{+0}$&
$2.2804\cdot10^{-1}$&
$2.8397\cdot10^{-1}$
\tabularnewline
\hline
$1030$&
$3.1537\cdot10^{-5}$&
$8.5154\cdot10^{-1}$&
$2.4661\cdot10^{+0}$&
$2.4661\cdot10^{+0}$&
$2.1821\cdot10^{-1}$&
$2.7085\cdot10^{-1}$
\tabularnewline
\hline
$1045$&
$2.9334\cdot10^{-5}$&
$8.2026\cdot10^{-1}$&
$2.3642\cdot10^{+0}$&
$2.3642\cdot10^{+0}$&
$2.0966\cdot10^{-1}$&
$2.5938\cdot10^{-1}$
\tabularnewline
\hline
$1060$&
$2.7306\cdot10^{-5}$&
$7.9263\cdot10^{-1}$&
$2.2747\cdot10^{+0}$&
$2.2747\cdot10^{+0}$&
$2.0218\cdot10^{-1}$&
$2.4925\cdot10^{-1}$
\tabularnewline
\hline
$1165$&
$1.6888\cdot10^{-5}$&
$6.6409\cdot10^{-1}$&
$1.8626\cdot10^{+0}$&
$1.8626\cdot10^{+0}$&
$1.6826\cdot10^{-1}$&
$2.0214\cdot10^{-1}$
\tabularnewline
\hline
$1210$&
$1.3884\cdot10^{-5}$&
$6.2993\cdot10^{-1}$&
$1.7540\cdot10^{+0}$&
$1.7540\cdot10^{+0}$&
$1.5955\cdot10^{-1}$&
$1.8952\cdot10^{-1}$
\tabularnewline
\hline
$1250$&
$1.1718\cdot10^{-5}$&
$6.0571\cdot10^{-1}$&
$1.6769\cdot10^{+0}$&
$1.6769\cdot10^{+0}$&
$1.5347\cdot10^{-1}$&
$1.8049\cdot10^{-1}$
\tabularnewline
\hline
$1355$&
$7.6472\cdot10^{-6}$&
$5.6024\cdot10^{-1}$&
$1.5317\cdot10^{+0}$&
$1.5317\cdot10^{+0}$&
$1.4237\cdot10^{-1}$&
$1.6323\cdot10^{-1}$
\tabularnewline
\hline
$1425$&
$5.8293\cdot10^{-6}$&
$5.3951\cdot10^{-1}$&
$1.4647\cdot10^{+0}$&
$1.4647\cdot10^{+0}$&
$1.3751\cdot10^{-1}$&
$1.5509\cdot10^{-1}$
\tabularnewline
\hline
$1500$&
$4.4031\cdot10^{-6}$&
$5.2277\cdot10^{-1}$&
$1.4098\cdot10^{+0}$&
$1.4098\cdot10^{+0}$&
$1.3374\cdot10^{-1}$&
$1.4828\cdot10^{-1}$
\tabularnewline
\hline
\end{tabular}
\end{footnotesize}
\end{center}
\caption{Percentage differences at NNLO for $g_Z=0.2$}
\label{table07}
\end{sidewaystable}

\begin{table}
\begin{center}
\begin{footnotesize}
\begin{tabular}{|c||c|c|c|c|c|c|}
\hline
\multicolumn{7}{|c|}{$d\sigma^{nnlo}/dQ$ [pb/GeV] for the FF model with $M_{Z^{\prime}}=2.5$ TeV, $\tan\beta=40$, Candia evol.}
\tabularnewline
\hline
$Q ~[\textrm{TeV}]$      &
$g_z=0.1$                 &
$g_z=0.4$                 &
$g_z=0.6$                 &
$g_z=0.8$                 &
$g_z=1$                   &
$\sigma_{nnlo}^{SM}(Q)$\tabularnewline
\hline
\hline
$2.400$&
$2.6475\cdot10^{-7}$&
$3.3941\cdot10^{-7}$&
$5.0947\cdot10^{-7}$&
$8.7995\cdot10^{-7}$&
$1.5720\cdot10^{-6}$&
$2.6141\cdot10^{-7}$
\tabularnewline
\hline
$2.423$&
$2.4961\cdot10^{-7}$&
$3.5212\cdot10^{-7}$&
$6.0291\cdot10^{-7}$&
$1.1654\cdot10^{-6}$&
$2.2223\cdot10^{-6}$&
$2.4543\cdot10^{-7}$
\tabularnewline
\hline
$2.446$&
$2.3629\cdot10^{-7}$&
$4.0068\cdot10^{-7}$&
$8.4077\cdot10^{-7}$&
$1.8529\cdot10^{-6}$&
$3.7317\cdot10^{-6}$&
$2.3050\cdot10^{-7}$
\tabularnewline
\hline
$2.469$&
$2.2656\cdot10^{-7}$&
$6.0047\cdot10^{-7}$&
$1.7162\cdot10^{-6}$&
$4.2536\cdot10^{-6}$&
$8.5322\cdot10^{-6}$&
$2.1654\cdot10^{-7}$
\tabularnewline
\hline
$2.492$&
$2.4932\cdot10^{-7}$&
$3.7446\cdot10^{-6}$&
$1.2697\cdot10^{-5}$&
$2.3281\cdot10^{-5}$&
$3.0409\cdot10^{-5}$&
$2.0349\cdot10^{-7}$
\tabularnewline
\hline
$2.5000$&
$3.7947\cdot10^{-5}$&
$3.7947\cdot10^{-5}$&
$3.7947\cdot10^{-5}$&
$3.7947\cdot10^{-5}$&
$3.7947\cdot10^{-5}$&
$1.9900\cdot10^{-7}$
\tabularnewline
\hline
$2.5003$&
$8.5271\cdot10^{-6}$&
$3.7283\cdot10^{-5}$&
$3.7757\cdot10^{-5}$&
$3.7858\cdot10^{-5}$&
$3.7892\cdot10^{-5}$&
$1.9886\cdot10^{-7}$
\tabularnewline
\hline
$2.5005$&
$2.7949\cdot10^{-6}$&
$3.5983\cdot10^{-5}$&
$3.7438\cdot10^{-5}$&
$3.7730\cdot10^{-5}$&
$3.7824\cdot10^{-5}$&
$1.9873\cdot10^{-7}$
\tabularnewline
\hline
$2.5770$&
$1.5907\cdot10^{-7}$&
$1.4769\cdot10^{-7}$&
$2.2368\cdot10^{-7}$&
$5.0120\cdot10^{-7}$&
$1.1340\cdot10^{-6}$&
$1.6192\cdot10^{-7}$
\tabularnewline
\hline
$2.636$&
$1.3692\cdot10^{-7}$&
$1.2412\cdot10^{-7}$&
$1.3364\cdot10^{-7}$&
$1.9772\cdot10^{-7}$&
$3.6561\cdot10^{-7}$&
$1.3839\cdot10^{-7}$
\tabularnewline
\hline
$2.700$&
$1.1628\cdot10^{-7}$&
$1.0680\cdot10^{-7}$&
$1.0536\cdot10^{-7}$&
$1.2481\cdot10^{-7}$&
$1.8637\cdot10^{-7}$&
$1.1718\cdot10^{-7}$
\tabularnewline
\hline
\end{tabular}
\end{footnotesize}
\end{center}
\caption{NNLO cross sections for the FF model with a
$M_{Z^{\prime}}=2.5$ TeV for
values of the coupling constant $g_z$ larger than $g_z=0.1$}
\label{table08}
\end{table}

\end{document}